\documentclass[twocolumn,useAMS,usenatbib]{mn2e}
\usepackage{graphicx}
\usepackage{bm}

\newcommand{\rmd}{{\rm d}}
\newcommand{\rme}{{\rm e}}

\newcommand{\rmH}{{\rm H}}
\newcommand{\rmHe}{{\rm He}}

\newcommand{\hehp}{{{\rm HeH}^+}}

\newcommand{\bhm}{B(\rmH^-)}

\newcommand{\R}{{\cal R}}

\newcommand{\xHI}{x_{\rm HI}}
\newcommand{\xe}{x_e}
\newcommand{\x}[1]{x[{{#1}}]}
\newcommand{\n}{n}

\newcommand{\beq}{\begin{equation}}
\newcommand{\eeq}{\end{equation}}
\newcommand{\beqa}{\begin{eqnarray}}
\newcommand{\eeqa}{\end{eqnarray}}

\title[Cosmological production of $\rmH_2$]
{Cosmological production of H$_2$ before the formation of the first 
  galaxies}

\author[Hirata \& Padmanabhan]
{Christopher M. Hirata$^1$ and Nikhil Padmanabhan$^2$
\newauthor
\\$^1$School of Natural Sciences,
      Institute for Advanced Study, Einstein Drive,
      Princeton, NJ 08540, USA
\\$^2$Joseph Henry Laboratories, Department of Physics, Jadwin Hall, 
      Princeton University, Princeton, NJ 08544, USA
}

\date{\today}

\begin{document}
\maketitle

\begin{abstract}
Previous calculations of the pregalactic chemistry have found that a small 
amount of H$_2$, $\x{{\rm H}_2}\equiv n[\rmH_2]/n[\rmH]
\approx 2.6\times 10^{-6}$, is produced 
catalytically through the H$^-$, H$_2^+$, and HeH$^+$ mechanisms.  We 
revisit this standard calculation taking into account the effects of the 
nonthermal radiation background produced by cosmic hydrogen recombination, 
which is particularly effective at destroying H$^-$ via photodetachment.  
We also take into consideration the non-equilibrium level populations of 
H$_2^+$, which occur since transitions among the rotational-vibrational 
levels are slow compared to photodissociation.  The new calculation 
predicts a final H$_2$ abundance of $\x{{\rm H}_2}\approx 6\times 10^{-7}$ 
for the standard cosmology.  This production is due almost entirely to the 
H$^-$ mechanism, with $\sim 1$ per cent coming from HeH$^+$ and $\sim 
0.004$ per cent from H$_2^+$. We evaluate the heating of the diffuse 
pregalactic gas from 
the chemical reactions that produce H$_2$ and from rotational transitions 
in H$_2$, and find them to be negligible.
\end{abstract}


\section{Introduction}

One of the key problems in cosmology is to understand the physical and 
chemical state of the baryonic matter in the Universe.  At high redshift, 
the baryonic matter was fully ionized and co-existed with a thermalized 
radiation field (the cosmic microwave background, or CMB).  By redshift 
$z\sim 10^3$, the Universe had expanded and cooled to $\sim 3000\,$K, at 
which point the ionized nuclei and free electrons of the primordial plasma 
combined to form neutral atoms. This cosmic recombination was first 
studied theoretically by \citet{1968ApJ...153....1P} and 
\citet{1968ZhETF..55..278Z}.  The observations of the acoustic peaks in 
the CMB $TT$ and $TE$ power spectra \citep{2001ApJ...561L...1L, 
2002ApJ...571..604N, 2002ApJ...568...38H, 2003ApJS..148..161K, 
2006astro.ph..3450P, 2006astro.ph..3451H} provide direct evidence that 
cosmic recombination happened, and that it occurred over a narrow range in 
redshift, in accordance with predictions.

As the Universe continued to expand and cool, the formation of molecules 
became thermodynamically favourable. Since hydrogen is most abundant, 
one would expect the most abundant molecule to be 
H$_2$. However, unlike atomic recombination, which occurs shortly after 
it becomes thermodynamically favourable and proceeds nearly to completion 
(e.g. \citealt{2000ApJS..128..407S}), cosmological formation of molecules 
is slow and freezes out with a final abundance,  
$\x{\rmH_2} \equiv n[\rmH_{2}]/n[\rmH] \ll 1$.

Despite their small abundance, molecules in the early universe have been 
investigated for several reasons. The first is that the primordial gas, 
mainly of hydrogen and helium atoms, lacks the low-lying excitations 
necessary for cooling and therefore star formation at low temperatures. On 
the other hand, molecules (which possess low-lying rotational excitations) 
could provide the cooling necessary to form the first stars 
\citep{1967Natur.216..976S}.  However, recent calculations indicate that 
the primordial H$_2$ abundance is far too small for this, and that the 
only H$_2$ important for cooling of early haloes is formed in collapsed 
haloes (e.g. \citealt{1997ApJ...474....1T}).  A second reason for studying 
H$_2$ production is that the heating of the gas, either via rotational 
transitions induced by the CMB or the chemical energy released by 
formation of the molecules, could affect the temperature of the 
pregalactic gas.  Here even a small effect could be important for 
proposals to study the absorption of the CMB by pregalactic gas in the 
H~{\sc i} 21-cm line \citep{2004PhRvL..92u1301L}. Finally, there is the 
(perhaps academic) motivation to understand the composition of the 
primordial gas as part of elucidating the standard cosmological model.

The first calculation of the primordial H$_2$ abundance was by 
\citet{1967Natur.216..976S}. Noting that the direct radiative association 
of two H atoms is forbidden, they proposed that H$_2$ molecules could be 
built up using H$_2^+$ as an intermediate state (the specific reactions 
will be given in Section~\ref{sec:rxn}).  \citet{1968ApJ...154..891P} and 
\citet{1969PThPh..41..835H} suggested that the H$^-$ mechanism dominated 
the production of molecules in primordial gas clouds.  A number of 
subsequent studies considered in increasing detail the H$_2$ abundance and 
cooling in primordial clouds \citep{1969PThPh..42..523H, 
1972PASJ...24...87Y, 1976ApJ...205..103H}.  \citet{1984ApJ...280..465L} 
performed a calculation of the abundances of H$_2$ as well as HD and LiH, 
calculating a final abundance $\x{\rmH_2}\sim 10^{-6}$ in the 
intergalactic gas. This is essentially today's ``standard'' calculation of 
the primordial H$_2$ abundance, although some of the reaction rates and 
cosmological parameters have been updated.  These updated analyses, which 
include substantial revisions to the deuterium and lithium chemistry, can 
be found in \citet{1993A&A...267..337P}, \citet{1995ApJ...451...44P}, 
\citet{1998A&A...335..403G}, \citet{1998ApJ...509....1S}, and references 
therein.

The effect of H$_2$ on heating of the gas before collapse has also been 
considered.  \citet{1993A&A...267..337P} presented the first analysis of 
the thermal effect of molecules; they found a moderate effect ($\sim$ 10 
per cent) on the gas temperature mainly due to rotational lines in H$_2$, 
and a smaller effect due to chemical reactions.  The heating in rotational 
lines was also considered by \citet{1996ApJ...464..523H} and 
\citet{1998A&A...335..403G}, but was revised downward by 
\citet{2000MNRAS.316..901F}, who concluded the effect was insignificant.  
We revisit the chemical heating here, and conclude that it dominates over 
H$_2$ rotational lines. However, even this effect is probably too small to 
be detected by 21-cm experiments in the forseeable future as it changes 
the gas temperature at the $\sim 10^{-4}$ level.

All of these analyses, however, have been based on several common 
assumptions.  One is the assumption of a purely thermal radiation field, 
which is not completely correct because of the line and continuum 
radiation emitted during hydrogen and helium recombination (e.g. 
\citealt{1993ASPC...51..548R, 2005astro.ph.10634W}).  Indeed, 
\citet{2005PhRvD..72h3002S} found that this spectral distortion suppresses 
the lithium abundance by several orders of magnitude as compared with 
previous calculations \citep{1996ApJ...458..401S, 1998A&A...335..403G, 
2002ApJ...580...29S}. The other assumption is the use of H$_2^+$ 
photodissociation rates based either on local thermodynamic equilibrium 
(LTE) populations of the rotational-vibrational levels, or with all 
H$_2^+$ ions in the ground state. It is however known that H$_2^+$ forms 
preferentially in excited states and since radiative transitions between 
levels are slow, LTE may not apply.  The importance of this was recognized 
by \citet{1998A&A...335..403G} and \citet{2002JPhB...35R..57L}, but a full 
non-LTE analysis of H$_2^+$ level populations has not been done.

Our purpose here is to revisit the calculation of H$_2$ abundance, 
including spectral distortions to the CMB and with a level-resolved 
treatment of H$_2^+$.  In Section~\ref{sec:rxn}, we introduce the chemical 
reactions important for H$_2$ production.  The H$^-$ mechanism, including 
the effect of the spectral distortion, is discussed in 
Section~\ref{sec:-}.  The H$_2^+$ mechanism and the level-resolved 
treatment is in Section~\ref{sec:+}, and HeH$^+$ is discussed in 
Section~\ref{sec:he}.  The abundances of H$_2$, H$^-$, H$_2^+$, and 
HeH$^+$ are calculated for our presently favoured cosmology in 
Section~\ref{sec:results}. The heating of the pregalactic gas by H$_2$ and 
the chemical reactions leading to its formation is considered in 
Section~\ref{sec:heating}.  We conclude in Section~\ref{sec:discussion}. 
The theory of the energy levels and transitions of the H$_2^+$ ion is 
recapitulated in Appendix~\ref{app:h2p}.

In this paper, we have assumed a primordial helium abundance of 
$Y_P=0.24$, and a flat $\Lambda$CDM cosmology with parameters from 
\citet{2005PhRvD..71j3515S}: $\Omega_b=0.0462$, $\Omega_m=0.281$, and 
$H_0=71.0\,$km$\,$s$^{-1}\,$Mpc$^{-1}$.  The number density $\n$ will 
refer to the total proper density of hydrogen nuclei in all forms 
(ionized, atomic, and molecular), although in the regime of interest here 
it is mostly atomic.  The notation $x_i$ (or $\x i$ for H$^-$, H$_2^+$, 
and HeH$^+$) will denote the number density of species $i$ relative to the 
total number density of hydrogen nuclei in all chemical forms (e.g. 
$\x{\rmH_2}= n[\rmH_2]/n=1/2$ if all hydrogen is molecular).

\section{The reactions}
\label{sec:rxn}

Due to the lack of a dipole moment, it is forbidden for two H atoms to 
combine radiatively to form H$_2$.  Therefore cosmological H$_2$ 
production proceeds through two main mechanisms catalyzed by charged 
particles \citep{1967Natur.216..976S, 1968ApJ...154..891P, 
1969PThPh..42..523H}.  The H$^-$ mechanism begins with
radiative attachment to form H$^-$,
\beq
\rmH + e^- \leftrightarrow \rmH^- + \gamma,
\label{eq:-1}
\eeq
and is completed when the associative detachment reaction
\beq
\rmH^- + \rmH \rightarrow \rmH_2 + e^-.
\label{eq:-2}
\eeq
A minor reaction that can use up H$^-$ ions is mutual neutralization,
\beq
\rmH^- + \rmH^+ \rightarrow 2\rmH.
\label{eq:-3}
\eeq

An alternative mechanism is via H$_2^+$, 
in which the catalyst is a proton rather than an electron,
\beq
\rmH + \rmH^+ \leftrightarrow \rmH_2^+ + \gamma.
\label{eq:+1}
\eeq
The H$_2^+$ ion can then be converted to H$_2$ via the reaction
\beq
\rmH_2^+ + \rmH \rightarrow \rmH_2 + \rmH^+.
\label{eq:+2}
\eeq
It is also possible for H$_2^+$ to be destroyed by dissociative 
recombination,
\beq
\rmH_2^+ + e^- \rightarrow 2\rmH.
\label{eq:+3}
\eeq

A third route is via the HeH$^+$ mechanism.  This is 
unimportant in the ``standard'' calculation, but given 
that we are revising the H$^-$ and H$_2^+$ rates downward it is only 
prudent to include it.  It begins with the production of HeH$^+$ by 
radiative association,
\beq
\rmHe+\rmH^+\leftrightarrow\rmHe\rmH^++\gamma.
\label{eq:he1}
\eeq
This ion could either be photodissociated (reverse of Eq.~\ref{eq:he1}), 
but it could also form H$_2^+$ by the reaction
\beq
\rmHe\rmH^+ + \rmH \leftrightarrow \rmHe + \rmH_2^+.
\label{eq:he2}
\eeq
The H$_2^+$ ion then participates in the usual sequence of reactions, 
Eqs.~(\ref{eq:+1}--\ref{eq:+3}).

All three of these mechanisms are suppressed at high redshift due to the 
intense CMB radiation, which drives Eqs.~(\ref{eq:-1}), (\ref{eq:+1}), and 
(\ref{eq:he1})
strongly to the left.  They are also inefficient at low redshift because 
the collisions required for them to proceed become rare as the universe 
expands and the gas density drops.  Past calculations have
H$_2$ production peaking at $z\sim 260$ for the H$_2^+$ mechanism and $z\sim 
90$ for the H$^-$ mechanism (e.g. \citealt{2002JPhB...35R..57L}).

Once formed, the H$_2$ molecule can be destroyed by UV photodissociation 
\citep{1967ApJ...149L..29S, 1992A&A...253..525A, 1996ApJ...467..522H}. 
Photoexcitation of H$_2$ from the ground 
$X^1\Sigma_g^+$ electronic state to either $B^1\Sigma_u^+$ (Lyman band 
absorption) or $C^1\Pi_u$ (Werner band absorption), is usually followed by 
radiative decay back to the $X^1\Sigma_g^+$ electronic state.  However, 
it is possible that this process leaves the H$_2$ 
molecule in an unbound vibrational state, resulting in its dissociation 
into two H atoms (a table of probabilities can be found in e.g. 
\citealt{1970ApJ...160L.107D}).  Of course, this relies on the 
existence of UV radiation in the Lyman and Werner bands, which have 
minimum energies of 11.2 and 12.3 eV respectively.  During the 
postrecombination era, these energies are 2--3 orders of magnitude larger 
than $kT_{\rm 
CMB}$ $\sim$ 0.2 eV, and so the photodissociation of H$_2$ plays 
a negligible role in determining the pregalactic H$_2$ abundance.  The 
presence of the spectral distortion does not change this situation since 
the distortion extends only up to 10.2 eV (the H~{\sc i} Ly$\alpha$ 
energy).  It is only after the first astrophysical sources turn on that 
intergalactic H$_2$ can be destroyed.

The cosmological production of H$_2$ can be followed by keeping track of 
the abundances of the relevant species: H, He, H$^+$, $e^-$, H$^-$, 
H$_2^+$, HeH$^+$, and H$_2$.  The evolution of H, He, H$^+$, and $e^-$ has 
been investigated in the context of the cosmic recombination, and is 
essentially unaffected by the rates of the catalytic reactions 
Eqs.~(\ref{eq:-1}--\ref{eq:+2}) due to the small amount of H$_2$ produced.  
We therefore use the recombination code {\sc Recfast} 
\citep{1999ApJ...523L...1S, 2000ApJS..128..407S} to compute 
the abundances of H, H$^+$, and 
$e^-$.  The rate equation for H$_2$ is
\beq
\frac{\rmd \x{\rmH_2}}{\rmd t} = S_{-} + S_{+},
\label{eq:H2}
\eeq
where $S_{-}$ and $S_{+}$ are the formation rates via the 
H$^-$ and combined H$_2^+$/HeH$^+$ mechanisms, respectively (note that 
the latter two cannot truly be separated since Eq.~\ref{eq:he2} couples
them).  In the next 
several sections, we compute the H$_2$ production and 
photodissociation rates.  Section~\ref{sec:-} treats the computation of 
$S_-$ 
(the H$^-$) mechanims.  Section~\ref{sec:+} treats the computation of 
$S_+$ 
including only the H$_2^+$ mechanism for simplicity.  In 
Section~\ref{sec:he} we 
introduce HeH$^+$ into the computation of $S_+$.

\section{The $\rmH^-$ mechanism}
\label{sec:-}

The rate for the H$^-$ mechanism depends on the population of H$^-$ as a 
function of time.  The relevant rate equation is
\beqa
\frac{\rmd\x{\rmH^-}}{\rmd t} &=& k_1\xe\xHI \n - k_{-1}\x{\rmH^-}
  - k_2\x{\rmH^-}\xHI\n
\nonumber \\ && - k_3\x{\rmH^-}\x{\rmH^+}\n,
\label{eq:H-}
\eeqa
where $k_1$ is the rate for Eq.~(\ref{eq:-1}), $k_{-1}$ is the rate for 
the reverse reaction, $k_2$ is the rate for Eq.~(\ref{eq:-2}), and $k_3$ 
is the rate for Eq.~(\ref{eq:-3}).  The forward rates are given by the 
fits in \citet{1998ApJ...509....1S}:
\beqa
k_1(T_m) &=& 3\times 10^{-16} \left(\frac{T_m}{300}\right)^{0.95}
      \rme^{-T_m/9320} \, {\rm cm}^3\,{\rm s}^{-1},
\nonumber \\
k_2(T_m) &=& 1.5\times 10^{-9} \left(\frac{T_m}{300}\right)^{-0.1}
      \, {\rm cm}^3\,{\rm s}^{-1}, {\rm ~and}
\nonumber \\
k_3(T_m) &=& 4\times 10^{-8} \left(\frac{T_m}{300}\right)^{-0.5}
      \, {\rm cm}^3\,{\rm s}^{-1},
\label{eq:k-rates}
\eeqa
where the matter temperature, $T_m$, is in Kelvin.  
(The stimulated radiative attachment can be 
neglected in comparison with the spontaneous rate because the energy of 
the emitted photon is always at least the H$^-$ binding energy, or 
$>0.754\,$eV; this is much greater than $k_BT_{CMB}$ in the redshift range of 
interest.)  There is some uncertainty in the associative detachment rate 
$k_2$ and the mutual neutralization rate $k_3$, which we will discuss in 
Section~\ref{sec:results}.
The ratio of H$^-$ destruction rate to the Hubble rate is always 
at least $k_2\n/H\approx 3500[(1+z)/10]^{1.3}\gg 1$, so we may treat 
$\x{\rmH^-}$ by the steady-state approximation,
\beq
\x{\rmH^-} = \frac{k_1\xe\xHI\n}
{k_2\xHI\n + k_{-1} + k_3\x{\rmH^+}\n}.
\label{eq:H-a}
\eeq
The production rate for H$_2$ via H$^-$ is then, 
\beq
S_{-} = \frac{k_1k_2\xe\xHI^2\n^2}
{k_2\xHI \n + k_{-1} + k_3\x{\rmH^+}\n}.
\label{eq:H-s}
\eeq

The photodetachment rate $k_{-1}$ depends on the details of the radiation 
field and can be broken into thermal (blackbody CMB) and nonthermal 
(spectral distortion) parts:
\beq
k_{-1} = k_{-1}^{({\rm th})}(T_{CMB}) + k_{-1}^{({\rm nt})}.
\eeq
The rate from the thermal photons can be computed via the principle of 
detailed balance,
\beqa
k_{-1}^{({\rm th})}(T_{CMB}) \!\! &=& \!\! 4 \left( 
\frac{m_ek_BT_{CMB}}{2\pi\hbar^2}
\right)^{3/2} e^{-\bhm/k_BT_{CMB}}
\nonumber \\ && \!\! \times
 k_1(T_{CMB}),
\label{eq:k-1th}
\eeqa
where the factor of 4 comes from the spin degeneracy (1 for H$^-$, 2 each
for H and $e^-$), and $\bhm=0.754\,$eV is the photodetachment
threshold energy.  Note that $T_{CMB}$ is used here instead of $T_m$ since
photodetachment depends only on the properties of the radiation field.

In the standard calculation the thermal rates are used, i.e. 
$k_{-1}=k_{-1}^{(th)}$. The nonthermal contribution to the
photodetachment rate is
\beq
k_{-1}^{({\rm nt})} = \n c \int_{\bhm/h}^\infty r(\nu)
  \sigma_{-1}(\nu)\, \frac{\rmd\nu}\nu,
\eeq
where $r(\nu)$ is the number of distortion photons per H atom per
logarithmic range in frequency, and $\sigma_{-1}(\nu)$ is the
photodetachment cross section, for which we use the fit by
\citet{1997ApJ...474....1T},
\beq
\sigma = 3.486 \times 10^{-16} \frac{(x-1)^{3/2}}{x^{3.11}} \,\,,
\eeq
where $x=h\nu/\bhm$. The variable $r(\nu)$ is related to the
phase space density $f(\nu)$ of photons by
\beq
f(\nu) = \frac{1}{\rme^{h\nu/k_BT_{CMB}}-1}
 + \frac{\n c^3}{8\pi\nu^3}r(\nu),
\eeq
where the first term denotes the thermal CMB contribution, and the latter 
term is the spectral distortion.  The spectral distortion
dominates at high frequencies, $h\nu/k_BT_{CMB}>30$.

The spectral distortion $r(\nu)$ is calculated as in 
\citet{2005PhRvD..72h3002S}: the H~{\sc i} $2s\rightarrow 1s$ two-photon 
decay and Lyman-$\alpha$ resonance escape rates were obtained from {\sc 
Recfast} and integrated as described in Section~II of 
\citet{2005PhRvD..72h3002S}.  We note that a recent computation by 
\citet{2005astro.ph.10634W} find an additional distortion due to He~{\sc 
i} $2^1S_0\rightarrow 1^1S_0$ two-photon decays and escape from the 
$2^1P^o_1\rightarrow 1^1S_0$ resonance.  As can be seen from Fig.~3 of 
\citet{2005astro.ph.10634W}, the He~{\sc i} distortion contributes 
significantly to the photon spectrum at short wavelengths 
$\lambda<140(1+z)^{-1}\,\mu$m, and could be a 
significant contribution to H$^-$ or H$_2^+$ destruction rates.  However, 
these photons are in the H~{\sc i} Lyman continuum ($\lambda<912\,$\AA) at 
$z\sim 1500$, when the universe is already optically thick ($\tau\sim 
10^7$) due to H~{\sc i} photoionization.  Thus we do not expect the 
high-energy spectral distortion from He~{\sc i} recombination to survive
and have not included it in our analysis.

\section{The $\rmH_2^+$ mechanism}
\label{sec:+}

The H$_2^+$ mechanism is more complicated to analyze than the H$^-$ 
mechanism because unlike H$^-$, the H$_2^+$ ion has many bound states.  
One must therefore determine the populations of each H$_2^+$ level, taking into 
account the radiative association rates to each level, the radiative and 
collisional rates for changing the rotational and vibrational quantum 
numbers, and the destruction rates by photodissociation and charge 
transfer. We start with an overview of 
the physics of the H$_2^+$ ion, and describe the full reaction network
that describes the ion. Finally, we show how the relevant physics can 
be captured by a ``two-level'' approximation; this provides a 
computationally simpler approach to the H$_2^+$ mechanism.

The H$_2^+$ ion is diatomic, and can be described by specifying 
the electronic state, a rotational quantum number $N$ describing the total 
orbital angular momentum, and a vibrational quantum number $v$ equal to 
the number of radial nodes.  The electronic states of interest to us are 
the ground state $1s\sigma_g$ ($X^2\Sigma_g^+$) and the first excited 
state $2p\sigma_u$ ($A^2\Sigma_u^+$); higher states are not accessible at 
the temperatures under consideration.  We ignore the 
spin-orbit coupling and hyperfine structure since their energy splittings 
are small compared to $k_BT_m$ or $k_BT_{CMB}$.  Thus the degeneracy of a 
given rotational-vibrational level is $g_{vN}=2(2N+1)g'_{\rm nuc}$, where 
the $2$ comes from the electron spin.  The nuclear degeneracy is forced by 
proton wave function antisymmetry considerations to be $g'_{\rm 
nuc}=1/4$ for spatially symmetric states and $g'_{\rm 
nuc}=3/4$ for spatially antisymmetric states.  Almost all of the 
bound states of H$_2^+$ are in the $1s\sigma_g$ ground electronic state, 
which has an attractive potential with a minimum energy of $E_{\rm 
min}=-2.79\,$eV at an internuclear separation $R=1.06\,$\AA.  The 
next-lowest electronic state $2p\sigma_u$ is repulsive (except for a weak 
attractive region at large distance due to the polarizability of H).

The radiative association reaction (Eq.~\ref{eq:+1}) begins with an 
H$(1s)$ atom and H$^+$ ion approaching each other.  This initial 
electronic state is a superposition of the $1s\sigma_g$ and $2p\sigma_u$ 
states of H$_2^+$.  Because of dipole selection rules, the system can only 
produce an H$_2^+(1s\sigma_g)$ ion from the $2p\sigma_u$ initial 
electronic state.  This state is repulsive, so at low initial energies the 
wave function is confined to large internuclear separation $R$.  Wave 
function overlap considerations then imply that radiative association to 
highly excited H$_2^+(1s\sigma_g)$ states is preferred, since these states 
have significant wave functions at large $R$.  Conversely, direct 
radiative association to the ground state ($N=v=0$) is suppressed.  
This state of H$_2^+$ can be populated by 
radiative transitions from excited states, but the inversion symmetry of 
the H$_2^+$ ion implies that these must be electric quadrupole 
transitions, hence they are slow, compared to 
destruction of H$_2^+$. This circumstance results in level populations of 
H$_2^+$ that are very far from LTE.  The non-LTE distribution, with 
higher-energy levels overpopulated relative to the Boltzmann distribution, 
results in photodissociation cross sections that are significantly higher 
than the commonly used LTE cross sections of \citet{1974JPhB....7.2025A} 
and \citet{1994ApJ...430..360S}.

\subsection{Rate equations}
\label{ss:network+}

The level populations of H$_2^+$ are determined by the solution of a 
network of production, destruction, and level-changing reactions.  
Schematically, one may write
\beq
\frac{\rmd x_i}{\rmd t} = s_i + \sum_j R_{ij}x_j - \sum_j R_{ji}x_i
  - \gamma_ix_i,
\label{eq:xi}
\eeq
where $s_i$ is the rate of production of the $i$th level of H$_2^+$ 
(in units of ions per H nucleus per second), $R_{ij}$ is the rate for 
transitions from the $i$th level to the $j$th level, and $\gamma_i$ is the 
rate for destruction of H$_2^+$ ions in the $i$th level.  Note that the 
level index $i$ encodes both the vibrational and rotational quantum 
numbers: $i=(v,N)$.  We track all 423 bound levels of the ground 
electronic state, which have quantum numbers ranging up to $v=19$ and 
$N=35$.

The source function of H$_2^+$ comes from the radiative association 
reaction (Eq.~\ref{eq:+1}) and its rate is given by
\beq
s_i = \alpha_i \x{\rmH^+}\xHI \n.
\label{eq:source}
\eeq
The level-changing rates include both radiative and collisional rates, 
$R_{ij}=R_{ij}^{({\rm rad})}+R_{ij}^{({\rm col})}$.  The radiative rates 
are given by the standard expression
\beq
R_{ij}^{({\rm rad})} = \left\{ 
\begin{array}{lcl} A_{ij}[1+f(\nu_{ij})] & & E(i)<E(j) \\
A_{ji}f(\nu_{ji}) g_i/g_j & & E(j)>E(i) \end{array} \right.,
\eeq
where $A_{ij}$ is the Einstein coefficient.  In principle there is also a 
collisional term $R_{ij}^{({\rm col})}$, which could accelerate H$_2$ 
production by de-activating H$_2^+$ ions into lower energy levels.  These 
lower-energy ions would survive longer since they suffer less 
photodissociation, and hence have a higher probability of undergoing 
charge transfer to produce H$_2$.  However, it is not possible to produce 
more H$_2$ molecules by this mechanism than there are de-activating 
H$_2^+$-H collisions.  We will show in Section~\ref{sec:results} that even
if we assume the Langevin rate for charge transfer H$_2^+$(H,H$^+$)H$_2$,
we find that only $\sim 0.005$ per cent of the H$_2$ is produced via the 
H$_2^+$ mechanism.  Thus our conclusions about the final H$_2$ abundance 
are unaffected except in the highly unlikely circumstance that the 
de-activation rate coefficient is several orders of magnitude larger than 
the Langevin rate.

The destruction of H$_2^+$ proceeds by photodissociation 
(reverse of Eq.~\ref{eq:+1}), charge transfer (Eq.~\ref{eq:+2}), or 
dissociative recombination (Eq.~\ref{eq:+3}), at a rate $\gamma_{i}$, 
\beq
\gamma_i = \beta_i  + k^{({\rm ct})}_i\xHI\n
+ k_i^{({\rm dr})}\xe \n.
\label{eq:gamma}
\eeq
The radiative association and dissociation rates $\alpha_i$ and 
$\beta_i$, 
and the quadrupole Einstein coefficients $A_{ij}$, are computed in 
Appendix~\ref{app:h2p}.  We examine the charge 
transfer (Section \ref{ss:ct}) and dissociative recombination 
(Section \ref{ss:dr}) reactions in the following sections.

As described in Section \ref{ss:dr}, we neglect $k_i^{({\rm dr})}$ since 
it is small during the regime where the H$_2^+$ mechanism is most active; 
but if it were included, its only possible effect would be to further 
reduce the already negligible H$_2$ yield from this mechanism.

Eq.~(\ref{eq:xi}) possesses a steady-state solution
\beq
{\bmath x} = {\mathbfss T}^{-1}{\bmath s},
\label{eq:xiss}
\eeq
where we have written the level populations as a vector and the matrix 
${\mathbfss T}$ is given by
\beq
T_{ij} = -R_{ij} + \delta_{ij}\left(\gamma_i + \sum_k R_{ki}\right),
\eeq
which is valid if all the eigenvalues of ${\mathbfss T}$ are large (fast) 
compared to the Hubble time.  The H$_2$ production rate via the H$_2^+$ 
mechanism is then
\beq
S_{+} = \n\xHI\sum_i k^{({\rm ct})}_ix_i.
\eeq

\subsection{Charge transfer}
\label{ss:ct}

In order to complete our analysis, we need the rate of the charge transfer 
reaction (Eq.~\ref{eq:+2}) as a function of the matter temperature for 
each level of the initial-state H$_2^+$ ion.  Unfortunately, there 
are no published computations of the state-resolved rates 
\citep{2002JPhB...35R..57L}.  Most pregalactic chemistry networks 
have used the value
\beq
k^{({\rm ct})} = (6.4\pm 1.2)\times 10^{-10}\, {\rm cm}^3\,{\rm s}^{-1}
\label{eq:karpas}
\eeq
measured by \citet{1979JChPh..70.2877K} in an ion cyclotron resonance 
device; however the dependence on the temperature and initial state was 
not determined.  \citet{2002PhRvA..66d2717K} has computed cross sections 
for Eq.~(\ref{eq:+2}) resolved into individual vibrational levels.
but was motivated by studies of controlled fusion plasmas and 
so only extends down to thermal energies ($\frac{3}{2}k_BT_m=0.1\,$eV at 
$z=300$) for some of the levels.

Given that the rate for Eq.~(\ref{eq:+2}) has not been measured or 
calculated accurately for all relevant levels, we have run our calculation 
for three different cases.  In case (A), we use the 
\citet{1979JChPh..70.2877K} rate coefficient (Eq.~\ref{eq:karpas}) for all 
levels; in case (B) we have used the Langevin rate coefficient, $k^{({\rm 
ct})} = 2.38\times 10^{-9}\,$cm$^3\,$s$^{-1}$; and in case (C) we have 
linearly interpolated the values of $\sigma v_{\rm rel}$ from 
\citet{2002PhRvA..66d2717K} and obtained a reaction rate by integrating 
over a Maxwellian energy distribution,
\beq
k^{({\rm ct})}_i = \frac{2}{\sqrt{\pi}\,(k_BT_m)^{3/2}}
\int_0^\infty E^{1/2} \rme^{-E/k_BT_m} \sigma v_{\rm rel} \rmd E,
\eeq
where $E$ is the kinetic energy of the H$+$H$_2^+$ system in the 
centre-of-mass frame and $v_{\rm rel}$ is the initial relative velocity of 
H and H$_2^+$. For energies less than the lowest tabulated value (0.17~eV 
in the case of the $v=0$ state) we have assumed $\sigma v_{\rm rel}$ to be 
constant.  This probably overestimates the reaction rate since for most 
levels $\sigma v_{\rm rel}$ is an increasing function of collision energy.  

\subsection{Dissociative recombination}
\label{ss:dr}

It is possible for dissociative recombination (Eq.~\ref{eq:+3}) to reduce 
the production of H$_2$.  In general dissociative recombination can be
important if it contributes significantly in Eq.~(\ref{eq:gamma}), i.e. if
\beq
k^{({\rm dr})}_i \ge \sim \frac{\xHI}{x_e} k^{({\rm ct})}_i
\approx x_e^{-1} k^{({\rm ct})}_i.
\label{eq:drcondition}
\eeq
Of the various models we consider, the lowest value of $k^{({\rm ct})}_i$ 
occurs for model A, with $k^{({\rm ct})}_i=6.4\times 
10^{-10}\,$cm$^3\,$s$^{-1}$.  At $z<600$ where the H$_2^+$ mechanism is 
most active, we have $x_e<10^{-3}$ so in order for dissociative 
recombination to be important, $k^{({\rm dr})}_i$ would have to be $\ge 
\sim 6\times 10^{-7}\,$cm$^3\,$s$^{-1}$.  The tabulated rate coefficients 
for the first 78 levels of H$_2^+$ are less than this for $20\le T_m\le 
5000\,$K (\citealt{1994ApJ...424..983S}; the maximum value in the table is 
$3.1\times 10^{-7}\,$cm$^3\,$s$^{-1}$ at $T_m=20\,$K, and $1.3\times 
10^{-7}\,$cm$^3\,$s$^{-1}$ for $T_m\ge 100\,$K); we therefore neglect 
dissociative recombination. This may not be a valid approximation at 
$z>600$ where $x_e>10^{-3}$, or for the 
highest excited levels of H$_2^+$ (for which no published rates are 
available).  If dissociative recombination is important in these 
circumstances, the effect would be to decrease the H$_2$ abundance at 
$z>300$.  There would not be a significant effect at lower redshifts, 
because at $z<300$ the production of H$_2$ is dominated by the H$^-$ 
and HeH$^+$ mechanisms; H$^-$ is unaffected by dissociative recombination, 
and HeH$^+$ produces H$_2^+$ in one of the low-lying states for which we 
have already concluded that dissociative recombination is irrelevant.

\subsection{A ``two-level'' approximation}
\label{sub:two_level}

Although the above discussion completely specifies the solution for the 
H$_2^+$ channel, it is useful to consider an approximate ``two-level'' 
solution. This serves both as a check of the more involved numerical 
calculations above, as well as highlighting the essential physics behind 
this channel.

We start by describing the H$_2^+$ channel by the following two level 
decomposition
\begin{itemize}
\item H$_2^+$ forms in an excited state H$_2^+(i)$, at a rate 
$\alpha_{i}$,
\begin{equation}
\label{eq:2level1}
\rmH + \rmH^{+} \rightarrow \rmH_2^+(i) \,\,,
\end{equation}
\item The excited ion is then either photodissociated at rate $\beta_i$,
\begin{equation}
\label{eq:2level-1}
\rmH_2^+(i) + \gamma \rightarrow \rmH + \rmH^+ \,\,,
\end{equation}
or transitions to a LTE distribution via a quadrupole transition at rate 
$\R_i$,
\begin{equation}
\label{eq:2level2}
\rmH_2^+(i) \rightarrow \rmH_2^+ \,\,.
\end{equation}
\end{itemize}
Note that although this is a two level description, we still consider all 
energy levels $i$; they are now however decoupled from each other.

Given Eqs.~\ref{eq:2level1} to \ref{eq:2level2}, we can write out the 
differential equations describing the time evolution of the abundances of 
these species. Furthermore, as before, Eqs.~\ref{eq:2level-1} and 
\ref{eq:2level2} occur significantly faster than the Hubble time, and so, 
we use the steady-state approximation to eliminate any explicit mention of 
the H$_2^+(i)$ abundance. This yields the following equation for the 
H$_2^+$ abundance,
\begin{equation}
\label{eq:2level_ode}
\frac{\rmd x[\rmH_2^+]}{\rmd t} = \left(\sum_{i}\frac{\alpha_{i} 
\R_i}{\R_{i}+\beta_i}\right) n\xHI x[\rmH^+] -
\beta x[\rmH_2^+] \,\,,
\end{equation}
where $\beta$ is now the LTE photodissociation rate
(determined by averaging $\beta_i$ as calculated in Appendix~\ref{app:h2p} 
over the Boltzmann distribution of levels at the CMB temperature). 
Finally, there remains the issue of what to 
assume for $\R_i$; for simplicity, we
assume that $\R_i$ is the total transition probability to lower levels,
\beq
\R_i = \sum_{E(j)<E(i)} A_{ji}[1+f(\nu_{ji})].
\eeq
Note that Eq.~\ref{eq:2level_ode} now
resembles the standard H$_2^+$ calculation with a suppressed radiative 
association rate,
\begin{equation}
\label{eq:2level_alpha1}
\alpha' = \sum_{i}\frac{\alpha_{i} \R_i}{\R_{i}+\beta_i}
\,\,.
\end{equation}
Note that the error in this approximation could go either direction.
This is because, although our nascent H$_2^+$ ion is likely to be produced 
in a highly excited state, we do not really know without the full 
multi-level calculation whether {\em after} it emits its first photon the 
resulting distribution is ``more excited'' or ``less
excited'' than Boltzmann.  We defer a comparison of this approximation 
with the full calculation 
until Section~\ref{sec:results}, and conclude by tabulating the suppressed 
association rate as a function of redshift (Table~\ref{tab:suppress_assoc}).

\begin{table}
\caption{\label{tab:suppress_assoc}The suppressed association rate as a 
function of redshift, Eq.~\ref{eq:2level_alpha1}).}
\begin{tabular}{rccrc}
\hline
$z\;\;$ & $\log_{10} \alpha'$ (cm$^{3}\,$s$^{-1}$) & &
$z\;\;$ & $\log_{10} \alpha'$ (cm$^{3}\,$s$^{-1}$) \\
\hline
$ 50$ & $-21.8824$ & &
$300$ & $-22.1023$ \\
$100$ & $-22.1048$ & &
$350$ & $-22.2281$ \\
$150$ & $-21.9867$ & &
$400$ & $-22.3786$ \\
$200$ & $-21.9676$ & &
$450$ & $-22.5421$ \\
$250$ & $-22.0103$ & &
$500$ & $-22.7089$ \\
\hline
\end{tabular}
\end{table}

\section{The $\rmHe\rmH^+$ mechanism}
\label{sec:he}

The HeH$^+$ mechanism is really an additional set of reactions that couple 
to Eqs.~(\ref{eq:+1}--\ref{eq:+3}).  Therefore the most straightforward 
way to include it is to add HeH$^+$ as an additional ``level'' in the 
network of Section~\ref{ss:network+}.  Due to its large dipole moment, 
HeH$^+$ 
has very short-lived excited levels (lifetime $\sim 10^{-3}\,$s; 
\citealt{1982ApJ...255..489R}).  These lifetimes are short compared to the 
photodissociation time from these levels or the time between collisions 
($\sim 10^8\,$s at $z=300$), so the distribution of level populations is 
determined entirely by the radiation field.  In the absence of a spectral 
distortion, then, it would be permissible to assume that the level 
populations of HeH$^+$ are in LTE at the CMB temperature $T_{CMB}$.

In reality there is a spectral distortion, however its effect on HeH$^+$ 
is negligible.  The distortion photons that would affect HeH$^+$ are those 
at $E>1.6[(1+z)/250]\,$eV, where the distortion is significant compared to 
the thermal CMB radiation; there is roughly 1 distortion photon in this 
energy range per H atom.  Since the rate coefficient for H$_2^+$ formation from HeH$^+$ 
via Eq.~(\ref{eq:he2}) is $9.1\times 
10^{-10}\,$cm$^3\,$s$^{-1}$ \citep{1979JChPh..70.2877K}, 
photodissociation of HeH$^+$ by distortion photons can be significant if 
the photodissociation cross section averaged over the HeH$^+$ level and 
distortion photon energy distributions is $\sigma c\sim 9.1\times 
10^{-10}\,$cm$^3\,$s$^{-1}$. The cross section for 
photodissociation from the $v=0$ level of HeH$^+$ is always $\sigma 
c<1.5\times 10^{-12}\,$cm$^3\,$s$^{-1}$ \citep{1982ApJ...255..489R}.  The 
cross section from the higher excited states $v=7$ and $v=8$ is 
$\sigma c\sim 3\times 10^{-9}\,$cm$^3\,$s$^{-1}$ near threshold 
\citep{1978JPhB...11.3349S} and could, in principle, be important, 
except that the fraction of HeH$^+$ in these states is $\ll 1$ (they lie 
1.56 and 1.62 eV above the ground state, as compared with 
$k_BT_{CMB}=0.07\,$eV at $z=300$).  Hence the HeH$^+$ levels are in LTE at the 
CMB temperature $T_{CMB}$ and HeH$^+$ can be followed as an ``effective 
1-level'' molecular ion.

It is straightforward to write down the additional terms in 
Eqs.~(\ref{eq:source}--\ref{eq:gamma}) to take 
into account the HeH$^+$ contribution.  The additional source is due to 
radiative attachment,
\beq
s_\hehp = \alpha_\hehp \x\rmHe \xHI \n,
\eeq
where $\alpha_\hehp$ is the radiative attachment rate coefficient.  The 
destruction term is simply the photodissociation rate,
\beq
\gamma_\hehp = \beta_\hehp.
\eeq
Finally the transition matrix ${\mathbfss R}$ picks up two additional 
terms due to Eq.~(\ref{eq:he2}).  One is the forward reaction term
\beq
R_{i,\hehp} = k_{\ref{eq:he2}} f^{(\ref{eq:he2})}_i \xHI \n,
\eeq
where $k_{\ref{eq:he2}}$ is the rate coefficient for Eq.~(\ref{eq:he2}) 
and $f^{(\ref{eq:he2})}_i$ is the branching fraction to the $i$th level 
of H$_2^+$.  The 
other is the reverse reaction, which can be obtained by the principle of 
detailed balance,
\beqa
R_{\hehp,i} &=& R_{i,\hehp} \rme^{[E(i)+D_0(\hehp)]/k_BT_m}
\nonumber \\&& \times
  \frac{2Q(\hehp,T_m)}{g_i}
 \left( \frac{\mu_{\hehp+\rmH}}
  {\mu_{\rmHe+\rmH_2^+}} \right)^{3/2}.
\eeqa
The ratio of reduced masses is 
$\mu_{\hehp+\rmH}/\mu_{\rmHe+\rmH_2^+}\approx 5/8$, and the partition 
function $Q(\hehp)$ is obtained at the matter temperature $T_m$ from 
\citet{2005MNRAS.357..471E}.  The binding energy is 
$D_0(\hehp)=1.84412\,$eV 
\citep{1998ApJ...508..151Z}.  The factor of 2 comes from the ground state 
degeneracy of H (He has degeneracy 1).  The remaining additional term in 
the transition matrix is $R_{\hehp,\hehp}=0$.

To compute the H$_2$ production, one must also know 
$\alpha_\hehp$, $\beta_\hehp$, $k_{(\ref{eq:he2})}$, and 
$f^{(\ref{eq:he2})}_i$.  For $\alpha_\hehp$, we have used the fit by 
\citet{1998A&A...335..403G} to the results of \citet{1982ApJ...255..489R}.  
This rate can in principle be increased by 
stimulated radiative association.  However \citet{1998ApJ...508..151Z} 
found that this increases the HeH$^+$ abundance by $<20$ per 
cent in the redshift range of interest, and since our analysis does not 
change the formation and/or destruction mechanisms a similar result would 
apply to our case.  We have thus not included a correction for stimulated 
radiative association (we will see that the HeH$^+$ mechanism is not a 
major source of primordial H$_2$, so a correction of this magnitude in 
HeH$^+$ abundance translates into a much smaller correction to the net 
H$_2$ production).  
For $\beta_\hehp$, we have used the principle of detailed balance,
\beqa
\beta_\hehp &=& \frac{1}{Q(\hehp,T_{CMB})} \left( \frac{\mu
k_BT_{CMB}}{2\pi\hbar^2} \right)^{3/2}
\nonumber \\ && \times
 \rme^{-D_0(\hehp)/k_BT_{CMB}} \alpha_\hehp(T_{CMB});
\eeqa
here $\mu$ is the reduced mass of He and H$^+$.  The rate coefficient 
$k_{(\ref{eq:he2})}$ has been measured by \citet{1979JChPh..70.2877K} to 
be $9.1\times 10^{-10}\,$cm$^3\,$s$^{-1}$.  For the 
$f^{(\ref{eq:he2})}_i$, which control the distribution of 
rotation-vibration levels in the final state of Eq.~(\ref{eq:he2}), there 
do not appear to be any published measurements or calculations.  We have 
thus considered two cases: one in which Eq.~(\ref{eq:he2}) populates all 
energetically available levels with their statistical ratios, and one in 
which the reaction always leaves H$_2^+$ in the ground state of either 
para-H$_2^+$ ($v=0$, $N=0$; probability 1/4) or ortho-H$_2^+$ ($v=0$, 
$N=1$; probability 3/4).  While the two cases lead to different level 
populations of H$_2^+$, the total abundances $\x{\rmH_2^+}(z)$ and 
$\x{\rmH_2}(z)$ are unaffected.  This is because, for any of the H$_2^+$ 
levels accessible to Eq.~(\ref{eq:he2}) at thermal energies, the most 
probable fate of the H$_2^+$ ion is to radiate away its vibrational energy 
on a timescale of order $10^7\,$s, and then to undergo charge exchange to 
produce H$_2$ on a timescale of order $10^9\,$s.  Thus the information 
about the initial $(v,N)$ distribution of the H$_2^+$ is erased.  
Photodissociation from the low-$v$ levels of H$_2^+$ is strongly 
suppressed due to lack of wave function overlap, and electric quadrupole 
excitation to high-$v$ levels followed by photodissociation is slower than 
charge transfer; thus photodissociation is not effective at depleting the 
H$_2^+$ produced by Eq.~(\ref{eq:he2}).  The case where all
energetically available levels are populated with their statistical ratios 
will be used in the rest of this paper.

\section{Results: $\rmH_2$ abundance}
\label{sec:results}

\begin{figure}
\includegraphics[angle=-90,width=3.2in]{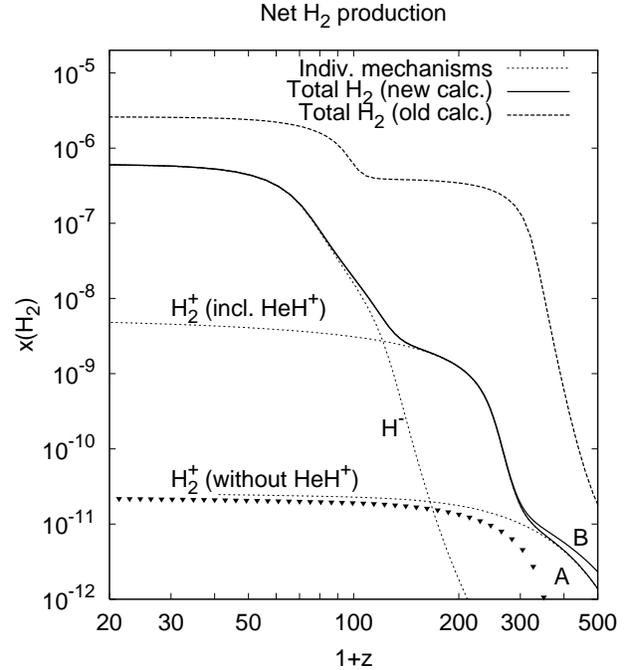}
\caption{\label{fig:H2_tot}The total abundance of H$_2$ as a function of 
redshift.  The thick solid lines shows the new calculation of H$_2$ 
abundance for models A and B (C is indistinguishable from B on the scale 
of the plot).  The thick long-dashed line shows the old calculation, which 
overestimated $x[\rmH_2]$.  The thin short-dashed lines show the 
individual production mechanisms: H$^-$, H$^+_2$/HeH$^+$, and H$^+_2$ 
(without HeH$^+$).  The triangles indicate the results for H$_2^+$ 
only from the simplified model of Section~\ref{sub:two_level}.}
\end{figure}

\begin{figure}
\includegraphics[angle=-90,width=3.2in]{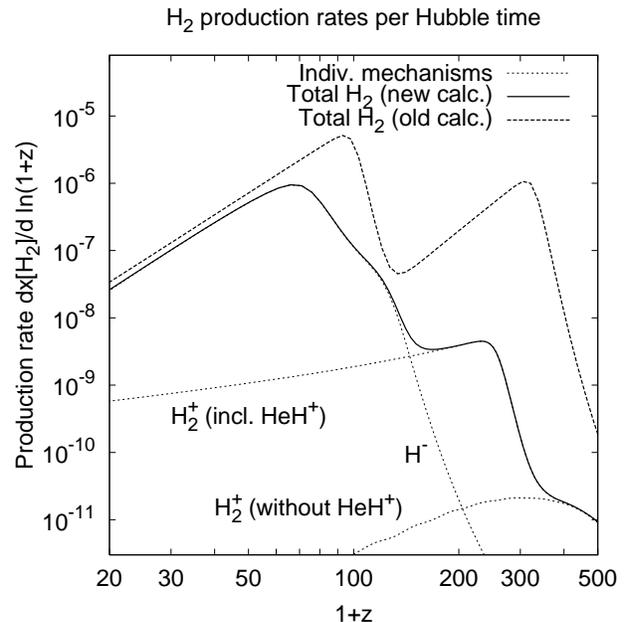}
\caption{\label{fig:shminus}The production rate of H$_2$ via the H$^-$, 
H$_2^+$, and HeH$^+$
mechanisms, in units of H$_2$ molecules per hydrogen nucleus per Hubble 
time.  The thick solid line shows the 
new calculation of H$_2$ rate for models A, while the thick 
long-dashed line shows the old calculation.  The thin short-dashed lines 
break down the H$_2$ production into the individual mechanisms.}
\end{figure}

\begin{figure}
\includegraphics[angle=-90,width=3.2in]{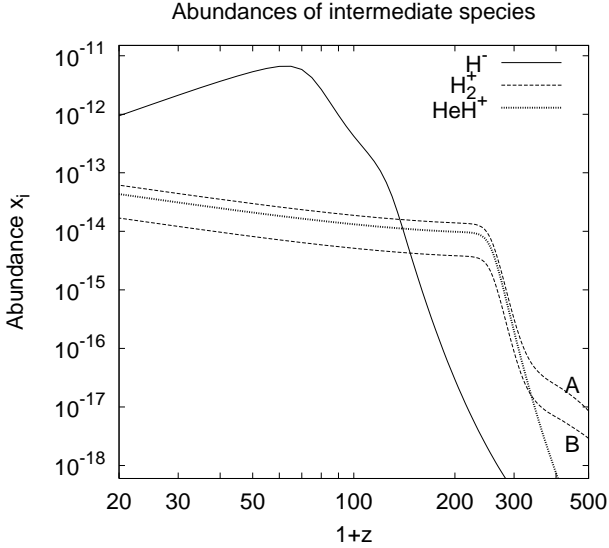}
\caption{\label{fig:hminus}The abundances of the intermediates H$^-$, 
H$^+_2$, and HeH$^+$.  The H$^+_2$ curve is shown for both models A and 
B.  Note that $x[$H$^-]$ peaks at $z\approx 62$ when H$_2$ production is 
also maximized.}
\end{figure}

We now present the results of integrating the production rate of H$_2$ 
from the three major mechanisms.  The total production of H$_2$ is 
obtained by the integral
\beq
\x{\rmH_2} = \int (S_{-}+S_+)\,\rmd t,
\eeq
which is shown in Fig.~\ref{fig:H2_tot}.  The production rate per 
Hubble time $(S_{-}+S_+)/H$ is shown in Fig.~\ref{fig:shminus}.  The 
abundances of the intermediates H$^-$, H$_2^+$, and HeH$^+$, obtained from 
Eqs.~(\ref{eq:H-a}, \ref{eq:xiss}), are shown in Fig.~\ref{fig:hminus}.

\subsection{H$^-$}

In our calculation, the final H$_2$ abundance is determined essentially 
entirely by the H$^-$ reaction sequence.  In accordance with previous 
calculations, at high redshift ($z>140$) the H$^-$ ion is formed mainly 
through radiative attachment, and destroyed mainly through photodetachment 
by thermal CMB photons.  Note that this is not quite a Saha-type 
equilibrium because of the different matter and radiation temperatures.  
At $z\approx 127$, the spectral distortion begins to dominate the photon 
spectrum at energies of $\sim 1\,$eV, where the H$^-$ photodetachment 
cross section peaks.  Below this redshift, H$^-$ is still produced mainly 
by radiative attachment, but the destruction mechanism is photodetachment 
from distortion photons.  This situation remains until $z\approx 67$, when 
most of the spectral distortion has redshifted to below the H$^-$ 
photodetachment threshold.  At this time, the competing photodetachment 
and associative detachment (Eq.~\ref{eq:-2}) rates are similar ($\sim 
10^{-10}\,$s$^{-1}$).  This is also the era of peak production of H$_2$: 
at higher redshifts the H$^-$ ions are destroyed before they can react 
with H to produce H$_2$, while at lower redshifts less H$^-$ is produced 
due to the lower density of the universe and the decrease in reaction rate 
at low $T_m$.  Ultimately we find that the total amount of H$_2$ produced 
via the H$^-$ mechanism, $\int S_-\rmd t$, is only $6\times 10^{-7}$ 
instead of $2.2\times 10^{-6}$ as found in the standard calculation.

As noted in Section~\ref{sec:-}, there is some uncertainty in the rate for 
associative detachment, Eq.~(\ref{eq:-2}).  \citet{2006ApJ...640..553G} 
argued that the rate coefficient $k_2$ could plausibly be varied between 
$6.5\times 10^{-10}$ and $5\times 10^{-9}\,$cm$^3\,$s$^{-1}$.  We have 
re-run our analysis using these values and find that the final H$_2$ 
abundance varies from $\x{\rmH_2}=4.1\times 10^{-7}$ for the lowest value 
of $k_2$ to $9.4\times 10^{-7}$ for the highest value.  Thus our fiducial 
estimate of $6\times 10^{-7}$ should be considered uncertain by a factor 
of $\sim 1.5$ in either direction.

There is also a large uncertainty in the mutual neutralization rate, with 
some results (e.g. the experimental work of \citealt{1970PhRvL..24..435M}) 
being up to an order of magnitude higher than the fits used here.  Using 
the higher mutual neutralization rate determined by the fit of 
\citet{1970PhRvL..24..435M}, we find that the final H$_2$ abundance 
decreases from $6.0\times 10^{-7}$ to $5.7\times 10^{-7}$.  Thus the 
uncertainty in the mutual neutralization rate does not have a significant 
effect on the final H$_2$ abundance.

\subsection{H$_2^+$ and HeH$^+$}

We find that the reactions involving positive ions 
(Eqs.~\ref{eq:+1}--\ref{eq:he2}) do not contribute significantly to the 
final H$_2$ abundance.  Nevertheless these reactions operate at earlier 
times than H$^-$ due to the higher binding energy of HeH$^+$ and H$_2^+$ 
compared to H$^-$, and so they dominate the H$_2$ production at $z>144$.  
The total amount of H$_2$ produced by these reactions is $\int S_+\rmd 
t=5\times 10^{-9}$, with almost all of this contributed by HeH$^+$.

\begin{figure}
\includegraphics[angle=-90,width=3.2in]{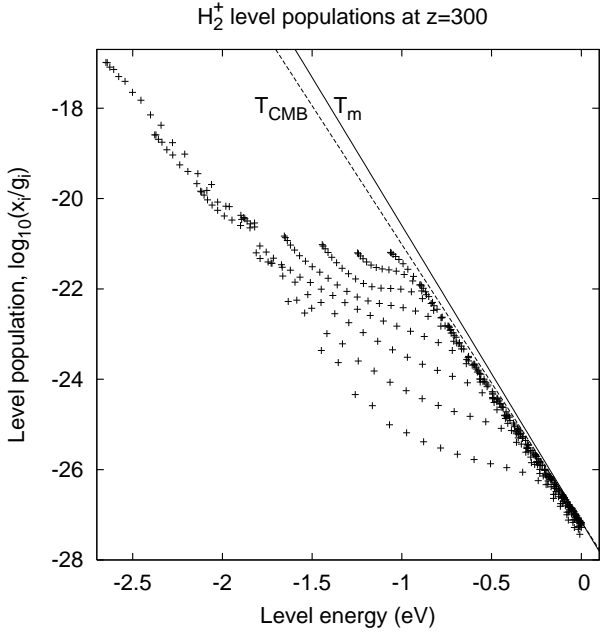}
\caption{\label{fig:h2p}The level populations in H$_2^+$ at $z=300$ for 
model A.  The 
horizontal axis shows the level energy with $E=0$ defined as the 
dissociation limit into H$+$H$^+$.  The actual populations for each of the 
423 rotational-vibrational levels are given by the crosses. The Boltzmann 
equilibrium populations $\propto \rme^{-E/k_BT}$ at the radiation and 
matter temperatures are shown by the straight lines.  Note that the levels 
close to the continuum are approximately in equilibrium, but the low-lying 
levels are highly underpopulated relative to Boltzmann.  The levels more 
tightly bound than $E(vN)<-1.844\,$eV have higher populations because they 
can be formed from HeH$^+$ via Eq.~(\ref{eq:he2}).  At this redshift the 
radiation and matter temperatures are $k_BT_{CMB} = 0.0708\,$eV and 
$k_BT_m = 0.0662\,$eV, respectively.}
\end{figure}

Our computed HeH$^+$ abundance (see Fig.~\ref{fig:hminus}) are very 
similar to those obtained in previous works \citep{1998A&A...335..403G, 
1998ApJ...509....1S}.  In contrast, we find much lower H$_2^+$ ion 
abundances (and H$_2$ production rates via H$_2^+$) than in the standard 
calculation.  This is a consequence of the non-Boltzmann level populations 
in H$_2^+$.  An example of these level populations for model A at 
$z=300$ is shown in 
Fig.~\ref{fig:h2p}.  (Results are qualitatively similar for model B, with 
the main quantitative difference being that the lowest-energy states are 
less populated.)  As shown in the figure, the highest-lying 
levels are nearly in Boltzmann equilibrium since the reaction
\beq
\rmH + \rmH^+ \leftrightarrow \rmH_2^+(1s\sigma_g,vN) + \gamma
\eeq
is fast compared with the electric quadrupole transitions in 
H$_2^+$.  (The slight deviation from Boltzmann equilibrium among the 
high-lying levels in the figure is due to the difference between matter 
and radiation temperatures.)  The populations of the lower-lying levels 
are determined by a combination of quadrupole radiative cascade rates, 
sourcing by HeH$^+$ through Eq.~(\ref{eq:he2}), and (at the lowest 
energies) some quadrupole excitation by the CMB.  Note that even in the 
ground vibrational state ($v=0$), the rotational levels of H$_2^+$ never 
come to thermal equilibrium with the CMB because the radiative rates 
(which may be as long as $\sim 10^{10}\,$s) are slower than the timescale 
for charge transfer (about $3\times 10^8\,$s for model A at $z=300$).

The overall production of H$_2$ via the HeH$^+$/H$_2^+$ mechanisms is 
found to be $\x{\rmH_2}=4.8\times 10^{-9}$ at $z=20$.  Of this, only a 
small fraction ($\x{\rmH_2}=2.6\times 10^{-11}$) is formed if we 
artificially turn off HeH$^+$.  (These numbers are for model A, models B 
and C give numbers that are $<10$ per cent higher.)  The direct production 
of H$_2^+$ via Eq.~(\ref{eq:+1}) only contributes significantly at $z>300$ 
where it is possible to directly produce tightly bound states by radiative 
association.

The two-level model of Section~\ref{sub:two_level} does quite well at 
reproducing the final H$_2$ abundance due to H$_2^+$ (it yields $2.2\times 
10^{-11}$ versus $2.6\times 10^{-11}$ for the full multi-level 
calculation).  It is not quite so good at reproducing the redshift 
history: at very high redshifts where H$_2^+$ can be excited out of the 
ground state and photodissociated, it underestimates H$_2$ production 
because the photodissociation rate of the H$_2^+$ ion after it has reached 
a low-lying level is actually less than the LTE rate.

\section{Heating}
\label{sec:heating}

The principal reason for interest in the molecular abundances in 
primordial gas is the possible effect on the heating and cooling rates.  
Before the formation of collapsed structures, the gas is colder than the 
CMB and hence the effect of molecular transitions is to heat the gas; 
after collapse the gas is compressed and shock-heated to above the CMB 
temperature, at which point molecules cool the gas.  The latter regime is 
dominated by H$_2$ molecules formed in the collapsed gas clouds themselves 
rather than primordial molecules, and hence it is not considered 
here.  The molecular heating in the pre-collapse regime was first 
investigated by \citet{1993A&A...267..337P}, and later by 
\citet{1996ApJ...464..523H}, \citet{1998A&A...335..403G}, and 
\citet{2000MNRAS.316..901F}.  The purpose of this section is to revisit 
the heating rate, taking into account the revised molecular production 
rates, and including the release of heat in chemical reactions.  The 
chemical heating was considered by \citet{1993A&A...267..337P} and found 
to be ``negligible,'' but no quantitative result was given; in any case we 
have found large changes in some of the relevant reaction rates and it is 
only prudent to re-evaluate the chemical heating term.

The overall temperature balance equation is
\beqa
\frac{\rmd T_m}{\rmd t} \!\!\! &=& \!\!\! -2HT_m + 
\frac{8\sigma_Ta_{bb}T_{CMB}^4}{3m_ec}(T_{CMB}-T_m)
\nonumber \\ && \!\!\!
+ \frac{2}{3k_Bx_t}(Q_{ff}+Q_{rec}+Q_{rot}+Q_c)
 - \frac{T_m}{x_t}\frac{\rmd x_t}{\rmd t},
\eeqa
where the first term represents the adiabatic expansion, the second term 
represents Compton heating, the third term represents heating by molecular 
rotation and chemical reactions, and the fourth term takes account of the 
changing number of translational degrees of freedom among which kinetic 
energy is distributed \citep{1993A&A...267..337P}.  In this equation, 
$\sigma_T$ is the Thomson cross section; $a_{bb}$ is the blackbody 
radiation coefficient; $m_e$ is the electron mass; $x_t$ is the number of 
gas particles per H nucleus; $Q_{ff}$ is the free-free heating 
rate; $Q_{rec}$ is the net heating due to 
recombinations (note that this is negative); $Q_{rot}$ is the heating by 
molecular rotation per H nucleus per unit time; and $Q_c$ is the heating 
by chemical reactions per H nucleus per unit time.

\subsection{Rotational transitions}

In this section we evaluate the heating in the rotational lines of H$_2$.  
It is found to be negligible, and is much smaller than the direct heating 
produced by the chemical reactions that generate H$_2$.  Determining the 
rotational line heating requires determining first the level populations 
and then computing the heating rates by considering all collisional 
excitations and de-excitations via H and H$^+$.

The H$_2$ molecule has two sets of levels with different nuclear spin 
properties, namely para-H$_2$ (even angular momentum $J$, and total 
nuclear spin $I=0$) and ortho-H$_2$ (odd $J$; $I=1$).\footnote{The orbital 
angular momentum $N$ and the orbital plus electron spin angular momentum 
$J$ are identical for the H$_2$ electronic ground state since the term 
symbol is $^1\Sigma^+_g$.} In the high-redshift universe, the timescale 
for the electric quadrupole radiative transitions that connect H$_2$ 
levels with the same nuclear spin is fast compared to the timescale 
for collisions.  For example, using the radiative rates of 
\citet{1977ApJS...35..281T} and the collisional rates described below, we 
find that the collisional-to-radiative transition rate ratio at $z=250$ is 
0.04 for $J=0$; this rate is even less for higher rotational levels or 
lower redshifts.  Therefore we may treat the H$_2$ molecule as an 
effective two-level system, tracking separately the abundances of 
para-H$_2$ and ortho-H$_2$.  Within each set of levels (even or odd $J$), 
the populations are assumed to rapidly thermalize to the CMB temperature, 
$x[{\rm H}_2,J]\propto g_J\rme^{-E_J/k_BT_{CMB}}$.  Here the degeneracy is 
$g_J=(2J+1)/4$ for even $J$ or $g_J=3(2J+1)/4$ for odd $J$.  The evolution 
equation for the fraction ${\cal F}$ of the H$_2$ molecules in the ortho
form is
\beq
\frac{\rmd\cal F}{\rmd t} = ({\cal F}_{\rm prod}-{\cal F})
\frac\rmd{\rmd t}\ln x[\rmH_2]
+ \gamma_{p\rightarrow o}(1-{\cal F})
- \gamma_{o\rightarrow p}{\cal F},
\eeq
where ${\cal F}_{\rm prod}$ is the fraction of H$_2$ that is produced in 
the ortho levels; $\gamma_{p\rightarrow o}$ is the para-to-ortho 
transition rate (in s$^{-1}$); and $\gamma_{o\rightarrow p}$ is the 
ortho-to-para transition rate.  The transition rates are
\beq
\gamma_{p\rightarrow o} = n\left( x[\rmH^+]\langle\sigma_{p\rightarrow o} 
v\rangle_{\rmH_2+\rmH^+}
+ \xHI\langle\sigma_{p\rightarrow o} v\rangle_{\rmH_2+\rmH} \right),
\eeq
where the averages are taken over both thermal velocity at temperature 
$T_m$ and the Boltzmann 
distribution of H$_2$ rotational levels.  A similar equation holds for 
$\gamma_{o\rightarrow p}$.  We have taken the cross sections for 
H$_2$+H$^+$ from \citet{1990JChPh..92.2377G}.  For H$_2$+H, we have used 
the fits by \citet{1997MNRAS.288..627F}, with the correction for reactive 
scattering by \citet{1999MNRAS.305..802L}.
Also, since H$_2$ is usually produced in a highly excited
rotational state when it forms from H$^-$+H \citep{1991A&A...252..842L}, 
we have assumed here that ${\cal F}_{\rm 
prod}=3/4$ as would be suggested by the nuclear spin degeneracy.
Once the level populations are established, the heating rate is computed 
by assuming that an amount of energy $E_J-E_{J'}$ is added to the 
translational degrees of freedom of the gas for 
each collisional $J\rightarrow J'$ transition:
\beqa
Q_{rot} \!\! &=& \!\! \sum_{J,J'} \x{\rmH_2,J}(E_J-E_{J'})n
\bigl(
x[\rmH^+]\langle\sigma_{J\rightarrow J'}
v\rangle_{\rmH_2+\rmH^+}
\nonumber \\ && \!\!
+ \xHI\langle\sigma_{J\rightarrow J'}
v\rangle_{\rmH_2+\rmH}
 \bigr).
\eeqa
Our treatment of H$_2$ cuts off the rotational levels at $J_{\rm max}=9$, 
the highest level for which \citet{1990JChPh..92.2377G} provides 
H$_2$+H$^+$ cross sections; but we have found that at $z<400$ there is a 
$<10$ per cent change in the heating rate if we cut off the levels at 
$J_{\rm max}=7$ instead.

\subsection{Chemical heating}

We next consider the chemical heating from each of 
Eqs.~(\ref{eq:-1}--\ref{eq:he2}).  The first contribution comes from the 
formation and destruction of H$^-$, Eq.~(\ref{eq:-1}).  The rate of loss 
of kinetic energy from the forward reaction is $k_1\bar E(T_m)x_e\xHI 
\n$, 
where $\bar E(T_m)$ is the mean energy of the incident electron 
participating in the reaction.  This mean 
energy is given by multiplying the Maxwell distribution against the 
radiative association cross section,
\beqa
\bar E
&=& \frac{\int_0^\infty E^{2}\sigma_{ra}(E)\rme^{-E/k_BT_m}\,\rmd E}
{\int_0^\infty E\sigma_{ra}(E)\rme^{-E/k_BT_m}\,\rmd E}
\nonumber \\ &=&
k_BT_m^2
\frac\rmd{\rmd T_m}
\ln\int_0^\infty E\sigma_{ra}(E)\rme^{-E/k_BT_m}\,\rmd E
\nonumber \\ &=&
k_BT_m^2\frac\rmd{\rmd T_m}\ln[T_m^{3/2}k_1(T_m)]
\nonumber \\ &=&
k_BT_m\left(\frac32+\frac{\rmd\ln k_1}{\rmd\ln T_m}\right),
\eeqa
where in the last line we have re-expressed the energy integral in terms 
of the thermally averaged rate coefficient.
By the principle of detailed balance, the 
average energy of the photodetached electrons from the thermal part of the 
reverse reaction is given by the same function, $\bar E(T_{CMB})$.  For the 
electrons that are photodetached by the spectral distortion, the kinetic 
energy input is $h\nu-\bhm$.  Thus the overall heating term 
associated with Eq.~(\ref{eq:-1}) is
\beqa
Q_{c1}\!\!\! &=& -\left(\frac32 + 
\frac{\rmd\ln k_1(T_m)}{\rmd\ln T_m}\right)k_BT_m k_1x_e\xHI \n
\nonumber \\ &&
 + \x{\rmH^-}\Bigl\{\left(\frac32 +
\frac{\rmd\ln k_1(T_{CMB})}{\rmd\ln T_{CMB}}\right)k_BT_{CMB} 
k_{-1}^{({\rm th})}
\nonumber \\ &&
 + \int_{\bhm/h}^\infty r(\nu)\sigma_{-1}(\nu)
 [h\nu-\bhm]
\frac{\rmd\nu}\nu
\Bigr\}.
\label{eq:qc1}
\eeqa

The second contribution is the heating due to associative detachment, 
Eq.~(\ref{eq:-2}).  As noted by \citet{1993A&A...267..337P}, this reaction 
is exothermic by 3.72~eV; however it is not correct to set the heating 
term from Eq.~(\ref{eq:-2}) equal to 3.72~eV times the reaction rate, 
because this energy yield is distributed among both the kinetic energy of 
the ejected electron (which couples to the gas temperature) and the 
rotational and vibrational degrees of freedom of the H$_2$ molecule (which 
are radiated away on timescales of $\sim 10^6\,$s, i.e. much shorter 
than the collision timescale).  In order to compute the heating term, one 
needs 
to know the mean excitation energy of the final-state H$_2$ molecule.  
This can be determined from the $(v,J)$-resolved cross sections for 
associative detachment of H and H$^-$, which were computed by 
\citet{1979ApJ...228..635B} at an initial state energy of 0.0129~eV (which 
is roughly the initial energy of interest); this yields a mean excitation 
energy of 2.81~eV.  By subtraction we estimate that 0.91~eV of energy is 
available to heat the gas.  Thus we set
\beq
Q_{c2} = (0.91\,{\rm eV})k_2\x{\rmH^-}\xHI \n.
\eeq

A third contribution to the heating rate is from mutual neutralization, 
Eq.~(\ref{eq:-3}).  The branching fraction for this rate is small, but it 
is strongly exothermic (12.84~eV) and thus provides a potentially large 
amount of energy.  The final state products are H atoms, which have no 
rotational or vibrational degrees of freedom.  They do however have 
electronic degrees of freedom: it is energetically possible for the final 
state to be H$(1s)$+H$(nl)$ for any $0\le l<n\le 4$.  At low energies, 
nearly all of the neutralizations go to $n=3$
\citep{1986JPhB...19L..31F}, so the energy released into translational 
degrees of freedom is 0.76~eV.  The excited H atom decays by emitting 
H$\alpha$, Lyman, and/or 2-photon continuum radiation, none of which can 
heat the gas.  Therefore the mutual neutralization heating term is
\beq
Q_{c3} = (0.76\,{\rm eV})k_3\x{\rmH^-}\x{\rmH^+} \n.
\eeq

We also consider the heating and cooling from the H$_2^+$ and HeH$^+$ 
reactions.  The contribution from Eq.~(\ref{eq:+1}) is obtained by 
inserting the translational energy $E=E(vN)+h\nu$ in the integrals for 
radiative association (Eq.~\ref{eq:alphavn-f}) and photodissociation 
(Eq.~\ref{eq:bvn}).  The heating from the charge transfer reaction, 
Eq.~(\ref{eq:+2}), depends on the final rotational-vibrational state of 
the H$_2$ molecule, which is not resolved in our code.  We have assumed 
that the H$_2$ is produced in the ground state, which maximizes the 
heating from this reaction since it implies that the entire energy yield 
of the reaction is available to heat the gas; however Eq.~(\ref{eq:+2}) is 
not a significant source of chemical energy anyway.  The radiative 
association and dissociation of HeH$^+$ (Eq.~\ref{eq:he1}) was treated 
using the obvious analogue of Eq.~(\ref{eq:qc1}), without the spectral 
distortion since as argued earlier it is unimportant for HeH$^+$.  The 
contribution to the heating rate from all of the H$_2^+$ and HeH$^+$ 
reactions is negligible.

\subsection{Results}

\begin{figure}
\includegraphics[angle=-90,width=3.2in]{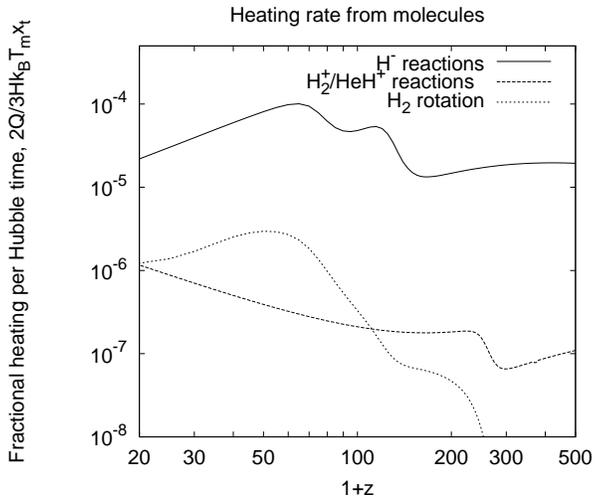}
\caption{\label{fig:Qmol}The heating rate due to the H$^-$ and 
H$_2^+$/HeH$^+$ sequences of reactions, and the H$_2$ rotational 
transitions, for Model A.  What is shown is the 
heating input per Hubble time, $Q/H$, divided by the thermal energy of 
the gas, $3k_BT_mx_t/2$.  The H$^-$ reactions 
affect the gas temperature at the $\sim 10^{-4}$, and H$_2^+$ reactions 
have even less influence.  The H$_2$ rotational line heating is always 
small compared to the chemical reactions.}
\end{figure}

The results are shown in Fig.~\ref{fig:Qmol}; it is seen that the chemical 
heating is very small, of the order of $10^{-4}$.  The principal source of 
heating is the H$^-$ sequence of reactions, which provide a peak in the 
heating at $z\sim 120$ from the photodetachment process, and a second peak 
at $z\sim 70$ due to associative detachment.  There is even less heating 
at earlier times, and in any case heating before Compton freeze-out 
($z\sim 200$) will be erased.

We find that the heating due to rotational transitions in H$_2$ is 
negligible, in qualitative agreement with the results of 
\citet{2000MNRAS.316..901F}.  The fractional heating rate peaks at $\sim 
3\times 10^{-6}$ at $z\approx 50$, shortly after the peak in H$_2$ 
production rate (there is less H$_2$ at earlier times, and the decreasing 
density and temperature suppress collisional rates at later times).  The 
difference from some previous results (e.g. \citealt{1993A&A...267..337P}) 
is due partially to our reduced H$_2$ abundance but also due to reduced 
$J$-changing collision cross sections.  In particular, collisions with 
H$^+$ dominate the transfer of rotational energy into translational 
degrees of freedom.  While there remains considerable uncertainty in the 
potential for the H$_2$+H system as noted by several authors (e.g. 
\citealt{1998A&A...335..403G, 2000MNRAS.316..901F}), we find that the 
cross sections would have to be increased by a factor of 40 to make 
rotational heating contribute at the $10^{-5}$ level 
($2Q_r/3Hk_BT_m=10^{-5}$), and by a factor of 500 to make it contribute at 
the $10^{-4}$ level.  Thus the conclusion that rotational heating is 
negligible is relatively insensitive to the remaining uncertainty in 
H$_2$+H cross sections.

Chemical and H$_2$ rotational heating are clearly very small effects and 
the only proposed observational technique that could achieve the $10^{-4}$ 
level of accuracy, even in principle, is 21-cm tomography 
\citep{2004PhRvL..92u1301L}.  In practice the experimental challenges of 
detecting the cosmological 21-cm signal are substantial (see e.g. 
\citealt{2006ApJ...638...20B} for a recent review), and observing the 
pre-reionization epoch to such high accuracy should be considered a goal 
for the distant future.  Until such data are available the chemical and 
rotational heating of the pregalactic gas can be safely neglected.

\section{Discussion}
\label{sec:discussion}

We have reconsidered the production of H$_2$ molecules in the pregalactic 
medium.  In contrast to previous studies, we have included the spectral 
distortion in our analysis, and resolved all 423 rotational-vibrational 
levels of the H$_2^+$ ion.  We find that in the level-resolved analysis, 
the H$_2^+$ reaction pathway is greatly suppressed because newly formed 
H$_2^+$ ions are photodissociated before they can decay to the ground 
state or undergo charge transfer to become H$_2$ molecules.  We also find 
that the H$^-$ ion is easily destroyed by spectral distortion photons at 
$z>70$, so that the production of H$_2$ by this pathway is suppressed 
relative to the standard calculation.  We obtain a final H$_2$ abundance 
$\x{\rmH_2}=6\times 10^{-7}$ assuming standard cosmology.

Unfortunately, the primordial H$_2$ molecules will be very difficult to 
detect.  The main effect of H$_2$ on the thermal history of the gas 
actually comes from the formation process (via the H$^-$ sequence, 
Eqs.~\ref{eq:-1}, \ref{eq:-2}) rather than rotational lines; however the 
effect is only of the order of $10^{-4}$.  In principle the proposed 21-cm 
tomography of the pre-reionization Universe could reach the sensitivity at 
which primordial H$_2$ becomes important, since it is sensitive to 
the temperature of the gas and has many more than $(10^{-4})^{-2}\sim 
10^8$ modes.  However when assessing the prospects, it should be 
remembered that the high-redshift 21-cm signal has not yet been detected, 
and measurements at the $10^{-4}$ level are clearly very far in the 
future.

Aside from H$_2$, there are other molecules with rotational lines such as 
HD and LiH, which could conceivably have been formed in the early Universe 
and played a role in the thermal balance.  The treatment of these trace 
molecules is beyond the scope of this paper, but HD in particular may 
warrant further study as it has been found to be a significant heating 
source in some past works (e.g. \citealt{1993A&A...267..337P}).  Since the 
main route of formation of HD is via the reaction H$_2$(D$^+$,H$^+$)HD 
\citep{1998A&A...335..403G}, any analysis of HD must incorporate the 
revised H$_2$ calculation presented here.

Finally, one could ask whether the H$_2$ suppression mechanisms discussed 
here -- the spectral distortion and non-equilibrium populations in H$_2^+$ 
-- have a significant effect on H$_2$ cooling of protogalaxies.  In the 
case of the spectral distortion, we have seen that at mean density the 
photodetachment of H$^-$ becomes unimportant at low redshift, since H$^-$ 
ions undergo a chemical reaction (usually associative detachment) before 
being destroyed by radiation.  For example, the branching fraction for 
H$^-$ at mean density to be destroyed by radiation is 0.13 at $z=40$, 0.09 
at $z=30$, and 0.05 at $z=20$.  In overdense gas clouds at $z<40$ the 
collisional reactions are faster and we conclude that the spectral 
distortion should be negligible.  We have tried running our H$_2^+$ code 
for overdense conditions at low redshift (e.g. $z=20$, $\delta_b=10^4$, 
$T=10^3\,$K) and find that the H$_2^+$ levels are still far out of 
equilibrium, with the lowest levels underpopulated relative to the 
Boltzmann distribution (at $T_{CMB}$) by many orders of magnitude.  
However in such clouds H$^-$ is likely to be a more important source of 
molecules than H$_2^+$ (see e.g. \citealt{1997ApJ...474....1T}).  
Therefore we do not expect our changes in H$_2^+$ physics to have large 
consequences for the cooling of the first collapsed objects in the 
universe.

\section*{Acknowledgments}

C.H. is supported in part by NSF PHY-0503584 and by a grant-in-aid from 
the W. M. Keck Foundation.  We thank Eric Switzer and Uro\v{s} Seljak for 
useful discussions.

\appendix

\section{The $\rmH_2^+$ molecular ion}
\label{app:h2p}

The bound states of H$_2^+(1s\sigma_g)$ are solved within the 
Born-Oppenheimer approximation, which for a diatomic molecular ion gives
\beq
\Psi({\bmath r},{\bmath R}) = \chi({\bmath r}|{\bmath R})
  \frac{\Phi(R)}{R}Y_{NM_N}(\hat{\bmath R}),
\eeq
where ${\bmath r}$ is the electron position, ${\bmath R}$ is the 
internuclear separation vector, $\chi({\bmath r}|{\bmath R})$ is the 
$1s\sigma_g$ electronic wavefunction, $\Phi(R)$ is the radial 
wavefunction, and $\hat{\bmath R}$ is the unit vector in the direction of 
${\bmath R}$.  The radial wavefunction is a solution to the 
Schr\"{o}dinger eigenvalue equation,
\beq
-\frac{\hbar^2}{2\mu}\Phi''(R) + \left[
  V(R) + \frac{\hbar^2N(N+1)}{2R^2}
  \right]\Phi(R) = E\Phi(R),
\label{eq:schrodinger}
\eeq
where $E$ is the bound state energy and $\mu$ is the reduced mass of 
H$_2^+$.  The potential $V(R)$ is interpolated from the tabulated values 
of \citet{bates53} and \citet{1971AD......2..171M}.  The radiative 
transitions among the levels of H$_2^+(1s\sigma_g)$ are treated according 
to the method of \citet{1953PPSA...66..784B}, who give the spontaneous rates 
$A_{ji}$ in terms of the wave functions\footnote{Note that $P(R)$ in 
\citet{1953PPSA...66..784B} is equal to our $\Phi(R)/R$.} and the 
quadrupole moment $M(R)$ of the H$_2^+$ ion.  The quadrupole moment is 
interpolated from the computations of \citet{1979JChPh..71.5382P} for 
$R\le 12a_0$, and using the asymptotic formula 
$M(R)\sim R^2/4-\alpha_da_0^3R^{-1}$ for $R>12a_0$ (where $\alpha_d=9/2$ 
is the dipole polarizability of H).

The rate for the radiative association reaction (Eq.~\ref{eq:+1}) is 
required for our calculation.  The individual rates to and from each 
rotational-vibrational level of H$_2^+$ as a function of matter and 
radiation temperature are not tabulated in the literature, but can be 
derived using standard methods (e.g. \citealt{1990ApJ...365..239Z, 
1993ApJ...414..672S}).  For the dipole radiative dissociation, one solves 
for the unbound final state Eq.~(\ref{eq:schrodinger}) using the 
$2p\sigma_u$ potential and the energy normalization (see e.g. Eq.~4 of
\citealt{1990ApJ...365..239Z}).  Then the radiative dissociation cross 
section from the $vN$ bound level with energy $E(vN)<0$ is
\citep{1968PhRv..172....1D}
\beq
\sigma_{rd}^{E1} = \frac{8\pi^3\nu}{3c}\overline{M^2},
\eeq
where $h\nu$ is the photon energy and
\beqa
\overline{M^2} &=& (2N+1)^{-1} \Biggl[ N
\left| \int_0^\infty \Phi(R)D(R)f_{N-1}(R) \rmd R \right|^2
\nonumber \\ &&
+ (N+1) \left| \int_0^\infty \Phi(R)D(R)f_{N+1}(R) \rmd R \right|^2 
\Biggr].
\eeqa
Here $D(R)$ is the dipole matrix element between $1s\sigma_g$ and 
$2p\sigma_u$ states, and $f_{N\pm 1}(R)$ are the 
unbound wave functions that solve Eq.~(\ref{eq:schrodinger}) with 
rotational quantum numbers $N\pm 1$ and energy $E(vN)+h\nu$.  (Only 
unbound states where $N'=N\pm 1$ contribute due to dipole selection 
rules.)  The dipole moments are interpolated from \citet{bates51} and 
\citet{1973AD......5..167R} for $R<20a_0$, and the large-$R$ asymptotic 
expression \citep{1972JPhB....5.2175R} is used for $R\ge 20a_0$.

The net radiative dissociation rate, including both thermal and nonthermal 
contributions to the radiation field, is then given by integrating over 
all photon frequencies,
\beqa
\beta_{vN} &=& \frac{8\pi}{c^2}\int_{-E(vN)/h}^\infty 
\frac{\nu^2\sigma_{rd}(\nu)\,\rmd\nu}{\rme^{h\nu/k_BT_{CMB}}-1}
\nonumber \\ &&
  + \n c\int_{-E(vN)/h}^\infty r(\nu)\sigma_{rd}(\nu) 
\frac{\rmd\nu}{\nu}.
\label{eq:bvn}
\eeqa

The spontaneous radiative association cross section is related to the 
radiative dissociation cross section by phase space factors,
\beq
\sigma_{ra,vN}(E) = \frac{\varpi_{\rmH_2^+(vN)+\gamma}}
{\varpi_{\rmH+\rmH^+}}
\sigma_{rd,vN}[E-E(vN)],
\label{eq:sigra}
\eeq
where $\varpi$ represents the number of quantum states per unit 
wavenumber $k$ per unit volume for either $\rmH_2^+(vN)+\gamma$ or 
$\rmH+\rmH^+$.  These are given by
\beq
\varpi_{\rmH_2^+(vN)+\gamma} = 4(2N+1)g'_{\rm nuc}\frac{4\pi
k_\gamma^2}{(2\pi)^3}
\label{eq:v1}
\eeq
and
\beq
\varpi_{\rmH+\rmH^+} = 2\frac{4\pi k_{\rmH+\rmH^+}^2}{(2\pi)^3},
\label{eq:v2}
\eeq
where $k_\gamma=2\pi\nu/c$ is the photon wavenumber and 
$k_{\rmH+\rmH^+}=\sqrt{2\mu E}/\hbar$ is the relative wavenumber of H and 
H$^+$.  In addition to the usual factor $4\pi k^2/(2\pi)^3$, 
Eq.~(\ref{eq:v1}) contains a factor of 2 from the electron spin, a factor 
of 2 from the photon polarization, and the nuclear spin-degeneracy factor 
$g'_{\rm nuc}$; Eq.~(\ref{eq:v2}) contains a factor of 2 from the electron 
spin.  The net radiative association rate coefficient to the $vN$ level is 
then given by the usual integral of cross section times velocity over the 
Maxwell-Boltzmann distribution,
\beqa
\alpha_{vN} &=& \frac{2\sqrt{2}}{(\pi\mu)^{1/2}(k_BT_m)^{3/2}}
\nonumber \\ && \times \int_0^\infty
  E \sigma_{ra,vN}(E) \rme^{-E/k_BT_m} [1+f(\nu)] \,\rmd E.
\label{eq:alphavn}
\eeqa
Here $f(\nu)$ is the photon phase space density at frequency $\nu = 
[E-E(vN)]/h$, and it has been included to take into account stimulated 
radiative association.  The phase space density from spectral distortion 
photons is $\ll 1$ and can be neglected, so we appropximate 
$f(\nu)$ by a blackbody.  Substituting Eqs.~(\ref{eq:sigra}--\ref{eq:v2}) 
into Eq.~(\ref{eq:alphavn}) then gives the final result,
\beqa
\alpha_{vN} \!\! &=& \!\! \frac{2\sqrt{2}\,(2N+1)g'_{\rm nuc}h^3}
  {\pi^{1/2}c^2(\mu k_BT_m)^{3/2}}
\nonumber \\ && \!\! \times
  \int_{-E(vN)/h}^\infty \nu^2\sigma_{rd}(\nu)
  \frac{\rme^{[h\nu+E(vN)]/k_BT_m}}{1-\rme^{-h\nu/k_BT_{CMB}}} \rmd\nu.
\label{eq:alphavn-f}
\eeqa
In this equation $g'_{\rm nuc}=1/4$ for even $N$ and 
$3/4$ for odd $N$.


\begin{thebibliography}{}

\bibitem[\protect\citeauthoryear{{Abgrall} et~al.}{1992}]
  {1992A&A...253..525A}
{Abgrall} H., {Le Bourlot} J., {Pineau des For\^{e}ts} G., {Roueff} E., 
  {Flower} D.~R., {Heck} L., 1992, A\&A, 253, 525

\bibitem[\protect\citeauthoryear{{Argyros}}{1974}]
  {1974JPhB....7.2025A}
{Argyros} J.~D., 1974, J. Phys. B, 7, 2025

\bibitem[\protect\citeauthoryear{{Bates}}{1951}]
  {bates51}
{Bates} D.~R., 1951, J. Chem. Phys., 19, 1122

\bibitem[\protect\citeauthoryear{{Bates} \& {Poots}}{1953}]
  {1953PPSA...66..784B}
{Bates} D.~R., {Poots} G., 1953, Proc. Phys. Soc. A, 66, 784

\bibitem[\protect\citeauthoryear{{Bates} et~al.}{1953}]
  {bates53}
{Bates} D.~R., {Ledsham} K., {Stewart} A.~L., 1953, Phil. Trans. R. Soc.
  A, 246, 215

\bibitem[\protect\citeauthoryear{{Bieniek} \& {Dalgarno}}{1979}]
  {1979ApJ...228..635B}
{Bieniek} R.~J., {Dalgarno} A., 1979, ApJ, 228, 635

\bibitem[\protect\citeauthoryear{{Bowman} et~al.}{2006}]
  {2006ApJ...638...20B}
{Bowman} J.~D., {Morales} M.~F., {Hewitt} J.~N., 2006, ApJ, 638, 20

\bibitem[\protect\citeauthoryear{{Dalgarno} \& {Stephens}}{1970}]
  {1970ApJ...160L.107D}
{Dalgarno} A., {Stephens} T.~L., 1970, ApJ, 160, L107

\bibitem[\protect\citeauthoryear{{Dunn}}{1968}]
  {1968PhRv..172....1D}
{Dunn} G.~H., 1968, Phys. Rev., 172, 1

\bibitem[\protect\citeauthoryear{{Engel} et~al.}{2005}]
  {2005MNRAS.357..471E}
{Engel} E.~A., {Doss} N., {Harris} G.~J., {Tennyson} J., 2005, MNRAS, 357,
  471

\bibitem[\protect\citeauthoryear{{Flower}}{1997}]
  {1997MNRAS.288..627F}
{Flower} D.~R., 1997, MNRAS, 288, 627

\bibitem[\protect\citeauthoryear{{Flower} \& {Pineau des For\^ets}}{2000}]
  {2000MNRAS.316..901F}
{Flower} D.~R., {Pineau des For\^ets} G., 2000, MNRAS, 316, 901

\bibitem[\protect\citeauthoryear{{Fussen} \& {Kubach}}{1986}]
  {1986JPhB...19L..31F}
{Fussen} D., {Kubach} C., 1986, J. Phys. B, 19, L31

\bibitem[\protect\citeauthoryear{{Galli} \& {Palla}}{1998}]
  {1998A&A...335..403G}
{Galli} D., {Palla} F., 1998, A\&A, 335, 403

\bibitem[\protect\citeauthoryear{{Gerlich}}{1990}]
  {1990JChPh..92.2377G}
{Gerlich} D., 1990, J. Chem. Phys., 92, 2377

\bibitem[\protect\citeauthoryear{{Glover} et~al.}{2006}]
  {2006ApJ...640..553G}
{Glover} S.~C., {Savin} D.~W., {Jappsen} A.~K., 2006, ApJ, 640, 553

\bibitem[\protect\citeauthoryear{{Haiman} et~al.}{1996a}]
  {1996ApJ...464..523H}
{Haiman} Z., {Thoul} A.~A., {Loeb} A., 1996a, ApJ, 464, 523

\bibitem[\protect\citeauthoryear{{Haiman} et~al.}{1996b}]
  {1996ApJ...467..522H}
{Haiman} Z., {Rees} M.~J., {Loeb} A., 1996b, ApJ, 467, 522

\bibitem[\protect\citeauthoryear{{Halverson} et~al.}{2002}]
  {2002ApJ...568...38H}
{Halverson} N.~W. et~al., 2002, ApJ, 568, 38

\bibitem[\protect\citeauthoryear{{Hinshaw} et~al.}{2006}]
  {2006astro.ph..3451H}
{Hinshaw} G. et~al., 2006, preprint (astro-ph/0603451)

\bibitem[\protect\citeauthoryear{{Hirasawa}}{1969}]
  {1969PThPh..42..523H}
{Hirasawa} T., 1969, Prog. Theor. Phys., 42, 523

\bibitem[\protect\citeauthoryear{{Hirasawa} et~al.}{1969}]
  {1969PThPh..41..835H}
{Hirasawa} T., {Aizu} K., {Taketani} M., 1969, Prog. Theor. Phys., 41, 835

\bibitem[\protect\citeauthoryear{{Hutchins}}{1976}]
  {1976ApJ...205..103H}
{Hutchins} J.~B., 1976, ApJ, 205, 103

\bibitem[\protect\citeauthoryear{{Karpas} et~al.}{1979}]
  {1979JChPh..70.2877K}
{Karpas} Z., {Anicich} V., {Huntress} W.~T.~Jr., 1979, J. Chem. Phys., 70,
  2877

\bibitem[\protect\citeauthoryear{{Kogut} et~al.}{2003}]
  {2003ApJS..148..161K}
{Kogut} A. et~al., 2003, ApJS, 148, 161

\bibitem[\protect\citeauthoryear{{Krsti\'{c}}}{2002}]
  {2002PhRvA..66d2717K}
{Krsti\'{c}} P.~S., 2002, Phys. Rev. A, 66, 042717

\bibitem[\protect\citeauthoryear{{Launay} et~al.}{1991}]
  {1991A&A...252..842L}
{Launay} J.~M., {Le Dourneuf} M., {Zeippen} C.~J., 1991, A\&A, 252, 842

\bibitem[\protect\citeauthoryear{{Le Bourlot} et~al.}{1999}]
  {1999MNRAS.305..802L}
{Le Bourlot} J., {Pineau des For\^ets}, G., {Flower} D.~R., 1999, MNRAS,
  305, 802

\bibitem[\protect\citeauthoryear{{Lee} et~al.}{2001}]
  {2001ApJ...561L...1L}
{Lee} A.~T. et~al., 2001, ApJL, 561, L1

\bibitem[\protect\citeauthoryear{{Lepp} \& {Shull}}{1984}]
  {1984ApJ...280..465L}
{Lepp} S., {Shull} J.~M., 1984, ApJ, 280, 465

\bibitem[\protect\citeauthoryear{{Lepp} et~al.}{2002}]
  {2002JPhB...35R..57L}
{Lepp} S., {Stancil} P.~C., {Dalgarno} A., 2002, J. Phys. B, 35, R57

\bibitem[\protect\citeauthoryear{{Loeb} \& {Zaldarriaga}}{2004}]
  {2004PhRvL..92u1301L}
{Loeb} A., {Zaldarriaga} M., 2004, Phys. Rev. Lett., 92, 211301

\bibitem[\protect\citeauthoryear{{Madsen} \& {Peek}}{1971}]
  {1971AD......2..171M}
{Madsen} M.~M., {Peek} J.~M., 1971, Atomic Data, 2, 171

\bibitem[\protect\citeauthoryear{{Moseley} et~al.}{1970}]
  {1970PhRvL..24..435M}
{Moseley} J., {Aberth} W., {Peterson} J.~R., 1970, Phys. Rev. Lett., 24,
  435

\bibitem[\protect\citeauthoryear{{Netterfield} et~al.}{2002}]
  {2002ApJ...571..604N}
{Netterfield} C.~B. et~al., 2002, ApJ, 571, 604

\bibitem[\protect\citeauthoryear{{Page} et~al.}{2006}]
  {2006astro.ph..3450P}
{Page} L. et~al., 2006, preprint (astro-ph/0603450)

\bibitem[\protect\citeauthoryear{{Palla} et~al.}{1995}]
  {1995ApJ...451...44P}
{Palla} F., {Galli} D., {Silk} J., 1995, ApJ, 451, 44

\bibitem[\protect\citeauthoryear{{Peebles}}{1968}]
  {1968ApJ...153....1P}
{Peebles} P.~J.~E., 1968, ApJ, 153, 1

\bibitem[\protect\citeauthoryear{{Peebles} \& {Dicke}}{1968}]
  {1968ApJ...154..891P}
{Peebles} P.~J.~E., {Dicke} R.~H., 1968, ApJ, 154, 891

\bibitem[\protect\citeauthoryear{{Peek} et~al.}{1979}]
  {1979JChPh..71.5382P}
{Peek} J.~M., {Hashemi-Attar} A.-R., {Beckel} C.~L., 1979, J. Chem. Phys.,
  71, 5382

\bibitem[\protect\citeauthoryear{{Puy} et~al.}{1993}]
  {1993A&A...267..337P}
{Puy} D., {Alecian} G., {Le Bourlot} J., {L\'eorat} J.,
  {Pineau des For\^ets} G., 1993, A\&A, 267, 337

\bibitem[\protect\citeauthoryear{{Ramaker} \& {Peek}}{1972}]
  {1972JPhB....5.2175R}
{Ramaker} D.~E., {Peek} J.~M., 1972, J. Phys. B, 5, 2175

\bibitem[\protect\citeauthoryear{{Ramaker} \& {Peek}}{1973}]
  {1973AD......5..167R}
{Ramaker} D.~E., {Peek} J.~M., 1973, Atomic Data, 5, 167

\bibitem[\protect\citeauthoryear{{Roberge} \& {Dalgarno}}{1982}]
  {1982ApJ...255..489R}
{Roberge} W., {Dalgarno} A., 1982, ApJ, 255, 489

\bibitem[\protect\citeauthoryear{{Rybicki} \& {Dell'Antonio}}{1993}]
  {1993ASPC...51..548R}
{Rybicki} G.~B., {Dell'Antonio} I.~P., 1993, ASP Conf. Ser. 51, 548

\bibitem[\protect\citeauthoryear{{Saha} et~al.}{1978}]
  {1978JPhB...11.3349S}
{Saha} S., {Datta} K.~K., {Barua} A.~K., 1978, J. Phys. B, 11, 3349

\bibitem[\protect\citeauthoryear{{Saslaw} \& {Zipoy}}{1967}]
  {1967Natur.216..976S}
{Saslaw} W.~C., {Zipoy} D., 1967, Nature, 216, 976

\bibitem[\protect\citeauthoryear{{Schneider} et~al.}{1994}]
  {1994ApJ...424..983S}
{Schneider} I.~F., {Dulieu} O., {Giusti-Suzor} A., {Roueff} E., 1994, ApJ,
  424, 983

\bibitem[\protect\citeauthoryear{{Seager} et~al.}{1999}]
  {1999ApJ...523L...1S}
{Seager} S., {Sasselov} D.~D., {Scott} D., 1999, ApJL, 523, L1

\bibitem[\protect\citeauthoryear{{Seager} et~al.}{2000}]
  {2000ApJS..128..407S}
{Seager} S., {Sasselov} D.~D., {Scott} D., 2000, ApJS, 128, 407

\bibitem[\protect\citeauthoryear{{Seljak} et~al.}{2005}]
  {2005PhRvD..71j3515S}
{Seljak} U. et~al., 2005, Phys. Rev. D, 71, 103515

\bibitem[\protect\citeauthoryear{{Stancil} et~al.}{1993}]
  {1993ApJ...414..672S}
{Stancil} P.~C., {Babb} J.~F., {Dalgarno} A., 1993, ApJ, 414, 672

\bibitem[\protect\citeauthoryear{{Stancil}}{1994}]
  {1994ApJ...430..360S}
{Stancil} P.~C., 1994, ApJ, 430, 360

\bibitem[\protect\citeauthoryear{{Stancil} et~al.}{1996}]
  {1996ApJ...458..401S}
{Stancil} P.~C., {Lepp} S., {Dalgarno} A., 1996, ApJ, 458, 401

\bibitem[\protect\citeauthoryear{{Stancil} et~al.}{1998}]
  {1998ApJ...509....1S}
{Stancil} P.~C., {Lepp} S., {Dalgarno} A., 1998, ApJ, 509, 1

\bibitem[\protect\citeauthoryear{{Stancil} et~al.}{2002}]
  {2002ApJ...580...29S}
{Stancil} P.~C., {Loeb} A., {Zaldarriaga} M., {Dalgarno} A.,
  {Lepp} S., 2002, ApJ, 580, 29

\bibitem[\protect\citeauthoryear{{Stecher} \& {Williams}}{1967}]
  {1967ApJ...149L..29S}
{Stecher} T.~P., {Williams} D.~A., 1967, ApJ, 149, 29

\bibitem[\protect\citeauthoryear{{Switzer} \& {Hirata}}{2005}]
  {2005PhRvD..72h3002S}
{Switzer} E.~R., {Hirata} C.~M., 2005, Phys. Rev. D, 72, 083002

\bibitem[\protect\citeauthoryear{{Tegmark} et~al.}{1997}]
  {1997ApJ...474....1T}
{Tegmark} M., {Silk} J., {Rees} M.~J., {Blanchard} A., {Abel} T.,
  {Palla} F., 1997, ApJ, 474, 1

\bibitem[\protect\citeauthoryear{{Turner} et~al.}{1977}]
  {1977ApJS...35..281T}
{Turner} J., {Kirby-Docken} K., {Dalgarno} A., 1977, ApJS, 35, 281

\bibitem[\protect\citeauthoryear{{Wong} et~al.}{2006}]
  {2005astro.ph.10634W}
{Wong} W.~Y., {Seager} S., {Scott} D., 2006, MNRAS, 367, 1666

\bibitem[\protect\citeauthoryear{{Yoneyama}}{1972}]
  {1972PASJ...24...87Y}
{Yoneyama} T., 1972, Pub. Ast. Soc. Japan, 24, 87

\bibitem[\protect\citeauthoryear{{Zel'dovich} et~al.}{1968}]
  {1968ZhETF..55..278Z}
{Zel'dovich} Y.~B., {Kurt} V.~G., {Sunyaev} R.~A., 1968, Zh. Eksp. Teor.
  Fiz., 55, 278

\bibitem[\protect\citeauthoryear{{Zygelman} \& {Dalgarno}}{1990}]
  {1990ApJ...365..239Z}
{Zygelman} B., {Dalgarno} A., 1990, ApJ, 365, 239

\bibitem[\protect\citeauthoryear{{Zygelman} et~al.}{1998}]
  {1998ApJ...508..151Z}
{Zygelman} B., {Stancil} P.~C., {Dalgarno} A., 1998, ApJ, 508, 151

\end{thebibliography}
\end{document}